\documentclass[useAMS,usenatbib]{mnras}
\usepackage[T1]{fontenc}
\usepackage{ae,aecompl}
\usepackage{pdflscape}
\usepackage{graphicx}
\usepackage{longtable}
\usepackage{lscape}
\usepackage{amssymb}
\usepackage{endnotes} 
\usepackage{footnote}
\usepackage{subfigure}
\usepackage{amsmath}
\usepackage{txfonts}
\usepackage{natbib}
\usepackage{url}
\usepackage{breakurl}
\usepackage{helvet}
\usepackage[flushleft]{threeparttable}
\def\aj{{AJ}}
\def\araa{{ARA\&A}}
\def\apj{{ApJ}}
\def\apjl{{ApJ}}
\def\apjs{{ApJS}}
\def\aap{{A\&A}}
\def\aapr{{A\&A~Rev.}}
\def\mnras{{MNRAS}}
\def\pasp{{PASP}}
\title[ASASSN-16fp]{ASASSN-16fp (SN~2016coi): A transitional supernova between \\ Type Ic and broad-lined Ic}
\author[B. Kumar et al.]
{Brajesh Kumar$^{1}$\thanks{E-mail: brajesh.kumar@iiap.res.in}, 
{A. Singh}$^{1}$,
{S. Srivastav}$^{1}$, 
{D. K. Sahu}$^{1}$ and
{G. C. Anupama}$^{1}$ \\\\
$^{1}$Indian Institute of Astrophysics, II Block, Koramangala, Bangalore 560 034, India 
}

\date{Accepted ------------, Received ------------; in original form ------------}
\pagerange{\pageref{firstpage}--\pageref{lastpage}} \pubyear{}

\begin{document}
\label{firstpage}
\pagerange{\pageref{firstpage}--\pageref{lastpage}} \pubyear{2017}
\maketitle
\begin{abstract}

We present results based on a well sampled optical ($UBVRI$) and ultraviolet ({\it Swift}/UVOT) imaging, 
and low-resolution optical spectroscopic follow-up observations of the nearby Type Ic supernova (SN) 
ASASSN-16fp (SN~2016coi). The SN was monitored during the photospheric phase ($-10$ to $+$33 days with 
respect to the $B$-band maximum light). The rise to maximum light and early post-maximum decline of the 
light curves are slow. The peak absolute magnitude ($M_{V}$ = $-17.7\pm0.2$ mag) of ASASSN-16fp is 
comparable with broad-lined Ic SN~2002ap, SN~2012ap and transitional Ic SN~2004aw but 
considerably fainter than the GRB/XRF associated supernovae (e.g. SN~1998bw, 2006aj). 
Similar to the light curve, the spectral evolution is also slow. ASASSN-16fp shows distinct photospheric 
phase spectral lines along with the C\,{\sc ii} features. The expansion velocity of the ejecta near 
maximum light reached $\sim$16000 km s$^{-1}$ and settled to $\sim$8000 km s$^{-1}$, $\sim$1 month 
post-maximum. Analytical modelling of the quasi-bolometric light curve of ASASSN-16fp suggests that 
$\sim$0.1 M$_{\odot}$ $^{56}$Ni mass was synthesized in the explosion, with a kinetic energy
of 6.9$^{+1.5}_{-1.3}$ $\times$ 10$^{51}$ erg and total ejected mass of $\sim$4.5\,$\pm$\,0.3 M$_{\odot}$.
  
\end{abstract}


\section{Introduction}\label{intro}

A large fraction of massive stars undergo catastrophic explosions at the end point of their lives due to the 
gravitational collapse of their cores \citep[e.g.][]{1986ARA&A..24..205W,2003ApJ...591..288H,2009ARA&A..47...63S}. 
Such events are recognized as core-collapse supernovae (CCSNe). 
Various physical mechanisms play a crucial role during the evolutionary phases of these progenitor stars, and 
consequently, the post-explosion observational features (e.g light curve shapes, luminosities, spectral evolution 
etc.) exhibit heterogeneity. 
Among the larger group of CCSNe \citep[see][]{1941PASP...53..224M,1997ARA&A..35..309F} Type Ib, Ic and IIb constitute 
the `stripped-envelope, SE' \citep{1997ApJ...491..375C} sub-category where the progenitor's outer envelope of hydrogen
and/or helium is partially or completely removed before they explode. A radiation-driven stellar wind \citep*{2008A&ARv..16..209P}, 
eruptive mass loss \citep{2006ApJ...645L..45S} and/or mass transfer with a companion star \citep*{1985ApJ...294L..17W,
1992ApJ...391..246P,2010ApJ...725..940Y} are the likely mechanisms for the stripping of the outer envelopes. 
A sub-population of Type Ic (He poor or absent) SNe are differentiated with very broad absorption lines in their 
spectra which signify high expansion velocities of the ejecta ($\gtrsim$\,14000 km\,s$^{-1}$, a few days 
after the explosion). These events are commonly designated as `broad-line' Ic SNe (BL-Ic). A small fraction of 
them are linked with long-duration gamma ray bursts (GRBs) and X-ray Flash (XRF) e.g. SN~1998bw/GRB~980425 
\citep{1998Natur.395..670G,2001ApJ...555..900P} and SN~2006aj/XRF~060218 \citep{2006Natur.442.1011P}, while no such 
association is observed in a majority of the BL-Ic events e.g. 
SN~2012ap \citep{2015ApJ...799...51M},
SN~2009bb \citep{2011ApJ...728...14P},
SN~2007ru \citep{2009ApJ...697..676S},
SN~2003jd \citep{2008MNRAS.383.1485V},
SN~2002ap \citep{2002ApJ...572L..61M,2002MNRAS.332L..73G}
and SN~1997ef \citep{2000ApJ...534..660I}.  

In recent attempts, using large data samples, several groups have investigated the observational properties 
of different SE-SNe using single as well as multi-band photometry \citep[c.f.][]{2006AJ....131.2233R,
2014AJ....147..118R,2011ApJ...741...97D,2013MNRAS.434.1098C,2014ApJS..213...19B,2014ApJ...787..157P,
2015A&A...574A..60T,2016MNRAS.457..328L,2016MNRAS.458.2973P}. \citet{2001AJ....121.1648M} compared spectra of different 
SE-SNe and characterized individual SNe types in detail. More recently, \citet{2014AJ....147...99M} and 
\citet{2016APJ-637X-827-2-90} have, using an updated larger spectroscopic data set, provided with an identification 
scheme that helps constrain the progenitors of different kinds of SE-SNe. All these studies indicate that both Type Ic and 
BL-Ic SNe exhibit diversity in terms of their explosion parameters (e.g. mass of ejecta, mass of newly synthesized $^{56}$Ni 
and kinetic energy of the explosion). There are a few SE-SNe such as SN~2004aw that are known to have properties between 
Type Ic and BL-Ic SNe. These Ic events, termed as transitional Type Ic \citep{2006MNRAS.371.1459T}, are just a few in number. 
In order to understand them better, detailed studies of similar events are needed. The nearby supernova ASASSN-16fp exhibits 
properties similar to SN~2004aw and provides an opportunity to explore the observational properties of the transitional Ic events.

ASASSN-16fp was discovered \citep{2016ATel.9086....1H} in the outskirts of the nearby  galaxy UGC~11868 
(Type: Sm) on 2016 May 27.6 UT (JD~2457536.1) by the {\it Brutus} telescope of the All Sky Automated 
Survey for SuperNovae (ASAS-SN\footnote{\url{http://www.astronomy.ohio-state.edu/~assassin/index.shtml}}) 
project. The SN was located approximately 31.7 arcsec North and 7.9 arcsec West from the center of the 
host galaxy. The apparent brightness of the SN at the time of discovery was reported as $\sim$15.7 
mag ($V$-band). It was classified as a broad line Type Ic similar to SN~2006aj a few days before the 
maximum brightness \citep{2016ATel.9090....1E}.
Additional spectroscopic data obtained from various telescopes confirmed the broad-line nature of this object.
The absorption features at 5550 \AA\,, 6300 \AA\, and 6700 \AA\, present in the spectrum were identified as 
He\,{\sc i} 5876, 6678, and 7065 lines respectively, indicating possible presence of helium in the ejecta of 
ASASSN-16fp \citep{2016ATel.9124....1Y}.
This transient was also detected in the X-ray, UV \citep{2016ATel.9088....1G} and the radio domain 
\citep{2016ATel.9201....1N,2016ATel.9134....1M,2016ATel.9147....1A}. Table~\ref{tab_1} lists some of the basic 
parameters of ASASSN-16fp.     

We present in this paper the results of photometric (optical and UV) and low-resolution optical 
spectroscopic observations of ASASSN-16fp during the early phase. 


\begin{table}
\centering
\caption{Properties of ASASSN-16fp.}
\label{tab_1}
\begin{tabular}{llc}
\hline
Parameters       &     Value                                    &  Reference                 \\ \hline
RA (J2000)       & $\alpha$~=~21$^{\rm h}$~59$^{\rm m}$~04\fs14 &  1                         \\   
DEC (J2000)      & $\delta$~=~18\degr ~11\arcmin ~10\farcs46    &  1                         \\ 
Explosion epoch (UT)&     2016 May 25.9                         & Section~\ref{lc}           \\
                    &     (JD 2457534.4)                        &                            \\
Discovery date (UT) &      2016 May 27.6                        &  1                         \\
                    &     (JD 2457536.1)                        &                            \\
Distance            &    18.1 $\pm$ 1.3 Mpc                     &  2                         \\
$E(B-V)_{total}$    &    0.074 $\pm$ 0.002 mag                  &  Section~\ref{color}       \\ 
\hline
\end{tabular}  
\begin{minipage}{\textwidth}
$^{1}$ \citet{2016ATel.9086....1H}\\
$^{2}$ NASA/IPAC Extragalactic Database (NED) 
\end{minipage}
\end{table}


\begin{table*}
\centering
\caption{Identification number (ID) and calibrated magnitudes of stars in the field of ASASSN-16fp used 
as secondary standards. The quoted errors include both photometric and calibration errors.}
\label{tab_id}
\begin{tabular}{cccccc}
\hline
Star & $U$              & $B$              & $V$              & $R$              & $I$              \\
ID   & (mag)            & (mag)            & (mag)            & (mag)            & (mag)            \\ \hline
1    & 17.28 $\pm$ 0.03 & 17.18 $\pm$ 0.03 & 16.48 $\pm$ 0.02 & 16.10 $\pm$ 0.02 & 15.70 $\pm$ 0.02 \\
2    & 14.99 $\pm$ 0.02 & 15.03 $\pm$ 0.01 & 14.42 $\pm$ 0.01 & 14.07 $\pm$ 0.01 & 13.70 $\pm$ 0.01 \\
3    & 17.07 $\pm$ 0.02 & 16.47 $\pm$ 0.02 & 15.58 $\pm$ 0.01 & 15.10 $\pm$ 0.01 & 14.65 $\pm$ 0.01 \\
4    & 17.32 $\pm$ 0.03 & 16.47 $\pm$ 0.02 & 15.47 $\pm$ 0.01 & 14.94 $\pm$ 0.01 & 14.48 $\pm$ 0.01 \\
5    & 17.43 $\pm$ 0.03 & 16.69 $\pm$ 0.02 & 15.73 $\pm$ 0.01 & 15.19 $\pm$ 0.01 & 14.70 $\pm$ 0.01 \\
6    & 15.74 $\pm$ 0.02 & 15.63 $\pm$ 0.02 & 14.98 $\pm$ 0.01 & 14.60 $\pm$ 0.01 & 14.22 $\pm$ 0.01 \\
7    & 16.59 $\pm$ 0.02 & 16.07 $\pm$ 0.02 & 15.20 $\pm$ 0.01 & 14.72 $\pm$ 0.01 & 14.28 $\pm$ 0.01 \\
8    & 16.87 $\pm$ 0.02 & 16.61 $\pm$ 0.02 & 15.72 $\pm$ 0.01 & 15.20 $\pm$ 0.01 & 14.69 $\pm$ 0.01 \\
9    & 16.22 $\pm$ 0.02 & 16.08 $\pm$ 0.02 & 15.35 $\pm$ 0.01 & 14.92 $\pm$ 0.01 & 14.49 $\pm$ 0.01 \\
10   & 16.06 $\pm$ 0.02 & 15.96 $\pm$ 0.01 & 15.30 $\pm$ 0.01 & 14.90 $\pm$ 0.01 & 14.51 $\pm$ 0.01 \\
11   & 16.84 $\pm$ 0.02 & 16.86 $\pm$ 0.02 & 16.22 $\pm$ 0.02 & 15.82 $\pm$ 0.02 & 15.42 $\pm$ 0.02 \\
12   & 17.09 $\pm$ 0.03 & 16.59 $\pm$ 0.02 & 15.75 $\pm$ 0.01 & 15.29 $\pm$ 0.01 & 14.87 $\pm$ 0.01 \\
13   & 15.29 $\pm$ 0.02 & 14.95 $\pm$ 0.01 & 14.16 $\pm$ 0.01 & 13.72 $\pm$ 0.01 & 13.31 $\pm$ 0.01 \\
14   & 16.73 $\pm$ 0.02 & 16.58 $\pm$ 0.02 & 15.84 $\pm$ 0.02 & 15.42 $\pm$ 0.01 & 15.01 $\pm$ 0.01 \\
15   & 16.31 $\pm$ 0.02 & 16.24 $\pm$ 0.02 & 15.54 $\pm$ 0.01 & 15.12 $\pm$ 0.01 & 14.72 $\pm$ 0.01 \\
16   & 16.44 $\pm$ 0.02 & 16.46 $\pm$ 0.02 & 15.78 $\pm$ 0.01 & 15.36 $\pm$ 0.02 & 14.94 $\pm$ 0.01 \\
\hline
\end{tabular}
\end{table*}

\section{Observations and data reduction}\label{obs}

\begin{table*}
\centering
\caption{The $UBVRI$ photometric magnitudes of ASASSN-16fp. Error denotes 1$\sigma$ uncertainty.}
\label{tab_hct}
\begin{tabular}{cccccccc} 
\hline
Date        &  JD      & Phase$^{\ast}$    &  $U$             & $B$              & $V$              & $R$              & $I$     \\     
(yyyy-mm-dd)& 2457000+ &(d)       &  (mag)           & (mag)            & (mag)            & (mag)            & (mag)            \\  
\hline   
2016-05-29  & 538.4    & $-$9.8   & 15.37 $\pm$ 0.03 & 15.64 $\pm$ 0.03 & 14.95 $\pm$ 0.02 & 14.64 $\pm$ 0.02 & 14.69 $\pm$ 0.02 \\
2016-05-30  & 539.3    & $-$8.9   & 15.19 $\pm$ 0.03 & 15.46 $\pm$ 0.03 & 14.73 $\pm$ 0.02 & 14.46 $\pm$ 0.02 & 14.52 $\pm$ 0.02 \\
2016-05-31  & 540.4    & $-$7.8   & 15.07 $\pm$ 0.03 & 15.28 $\pm$ 0.03 & 14.51 $\pm$ 0.02 & 14.27 $\pm$ 0.02 & 14.35 $\pm$ 0.02 \\
2016-06-01  & 541.4    & $-$6.8   & 14.98 $\pm$ 0.02 & 15.15 $\pm$ 0.02 & 14.34 $\pm$ 0.02 & 14.14 $\pm$ 0.02 & 14.21 $\pm$ 0.02 \\
2016-06-02  & 542.5    & $-$5.8   & 14.92 $\pm$ 0.02 &  --              & 14.19 $\pm$ 0.02 & 14.01 $\pm$ 0.02 &  --              \\
2016-06-03  & 543.4    & $-$4.8   & 14.88 $\pm$ 0.02 & 14.93 $\pm$ 0.02 & 14.07 $\pm$ 0.02 & 13.87 $\pm$ 0.02 & 13.98 $\pm$ 0.02 \\
2016-06-05  & 545.4    & $-$2.8   & 14.86 $\pm$ 0.04 & 14.82 $\pm$ 0.04 & 13.90 $\pm$ 0.03 & 13.71 $\pm$ 0.03 & 13.83 $\pm$ 0.03 \\
2016-06-10  & 550.3    & $+$2.2   & 14.97 $\pm$ 0.02 & 14.80 $\pm$ 0.03 & 13.76 $\pm$ 0.02 & 13.53 $\pm$ 0.02 & 13.62 $\pm$ 0.02 \\
2016-06-11  & 551.4    & $+$3.3   &  --              &  --              &  --              & 13.52 $\pm$ 0.02 & --               \\
2016-06-13  & 553.4    & $+$5.2   & 15.14 $\pm$ 0.02 & 14.92 $\pm$ 0.03 & 13.77 $\pm$ 0.02 & 13.49 $\pm$ 0.02 & 13.54 $\pm$ 0.02 \\
2016-06-16  & 556.4    & $+$8.2   & 15.39 $\pm$ 0.02 & 15.13 $\pm$ 0.03 & 13.86 $\pm$ 0.02 & 13.52 $\pm$ 0.02 & 13.54 $\pm$ 0.02 \\
2016-06-18  & 558.4    & $+$10.2  & 15.62 $\pm$ 0.02 & 15.30 $\pm$ 0.03 & 13.95 $\pm$ 0.02 & 13.55 $\pm$ 0.02 & 13.56 $\pm$ 0.02 \\
2016-06-19  & 559.4    & $+$11.2  & 15.73 $\pm$ 0.02 & 15.39 $\pm$ 0.04 & 14.00 $\pm$ 0.02 & 13.58 $\pm$ 0.02 & 13.58 $\pm$ 0.03 \\
2016-06-21  & 561.3    & $+$13.2  & 15.98 $\pm$ 0.02 & 15.59 $\pm$ 0.02 & 14.12 $\pm$ 0.02 & 13.63 $\pm$ 0.02 & 13.62 $\pm$ 0.02 \\
2016-06-22  & 562.4    & $+$14.2  & 16.11 $\pm$ 0.04 & 15.68 $\pm$ 0.04 & 14.17 $\pm$ 0.03 & 13.67 $\pm$ 0.03 & 13.65 $\pm$ 0.03 \\
2016-06-24  & 564.5    & $+$16.3  &  --              &  --              & 14.28 $\pm$ 0.02 & 13.75 $\pm$ 0.02 & --               \\
2016-06-26  & 566.3    & $+$18.1  & 16.47 $\pm$ 0.02 & 15.96 $\pm$ 0.02 & 14.36 $\pm$ 0.02 & 13.82 $\pm$ 0.02 & 13.73 $\pm$ 0.02 \\
2016-06-30  & 570.4    & $+$22.2  & 16.82 $\pm$ 0.04 & 16.27 $\pm$ 0.03 & 14.64 $\pm$ 0.02 & 13.98 $\pm$ 0.03 & 13.87 $\pm$ 0.04 \\
2016-07-01  & 571.4    & $+$23.2  & 16.86 $\pm$ 0.03 & 16.31 $\pm$ 0.02 & 14.69 $\pm$ 0.02 & 14.05 $\pm$ 0.02 & 13.90 $\pm$ 0.02 \\
2016-07-03  & 573.5    & $+$25.3  &  --              &  --              & 14.79 $\pm$ 0.02 & 14.15 $\pm$ 0.02 & 13.97 $\pm$ 0.02 \\
2016-07-05  & 575.3    & $+$27.1  & 17.05 $\pm$ 0.02 & 16.47 $\pm$ 0.02 & 14.87 $\pm$ 0.03 & 14.20 $\pm$ 0.02 & 14.01 $\pm$ 0.02 \\
2016-07-06  & 576.2    & $+$28.1  & 17.08 $\pm$ 0.02 & 16.52 $\pm$ 0.02 & 14.93 $\pm$ 0.02 & 14.25 $\pm$ 0.02 & 14.03 $\pm$ 0.02 \\
2016-07-10  & 580.4    & $+$32.2  &  --              &  --              &  --              & 14.35 $\pm$ 0.04 & 14.12 $\pm$ 0.05 \\
2016-07-11  & 581.4    & $+$33.3  & 17.19 $\pm$ 0.03 & 16.65 $\pm$ 0.03 & 15.12 $\pm$ 0.02 & 14.42 $\pm$ 0.02 & 14.17 $\pm$ 0.03 \\
\hline
\end{tabular}\\
$^{\ast}$ With reference to the $B$-band maximum (JD~2457548.2).
\end{table*}

\begin{table*}
\centering
\caption{The {\it Swift}-UVOT photometric magnitudes of ASASSN-16fp, (in Vega-system).}
\small
\label{tab_swift}
\begin{tabular}{ccccccccc}
\hline
Date        &    JD     & Phase$^{\ast}$ & $u$              &  $b$              & $v$               & $uvm2$            & $uvw1$           &  $uvw2$          \\
(yyyy-mm-dd)& 2457000+  &  (d)           & (mag)            & (mag)             & (mag)             & (mag)             & (mag)            & (mag) \\  
\hline
2016-05-27  & 536.2     & $-$12.0        &15.88 $\pm$ 0.06  & 16.45 $\pm$ 0.07  & 15.78 $\pm$ 0.08  & 17.54 $\pm$ 0.13  & 16.53 $\pm$ 0.08 & 17.74 $\pm$ 0.16  \\
2016-05-29  & 537.9     & $-$10.3        &15.27 $\pm$ 0.03  & 15.81 $\pm$ 0.03  & 15.10 $\pm$ 0.03  & 17.00 $\pm$ 0.08  & 16.13 $\pm$ 0.05 & 17.09 $\pm$ 0.05  \\
2016-06-03  & 543.1     & $-$5.1         &14.69 $\pm$ 0.05  & 14.93 $\pm$ 0.05  & 13.97 $\pm$ 0.05  & 17.21 $\pm$ 0.11  &        --        & 16.77 $\pm$ 0.01  \\
2016-06-04  & 544.3     & $-$3.9         &14.69 $\pm$ 0.05  & 14.86 $\pm$ 0.04  & 13.86 $\pm$ 0.05  & 17.37 $\pm$ 0.12  & 15.96 $\pm$ 0.07 & 16.93 $\pm$ 0.09  \\
2016-06-05  & 544.5     & $-$3.7         &14.74 $\pm$ 0.05  & 14.91 $\pm$ 0.04  & 13.90 $\pm$ 0.04  & 17.26 $\pm$ 0.10  & 15.91 $\pm$ 0.07 & 17.07 $\pm$ 0.10  \\
2016-06-06  & 546.1     & $-$2.1         &14.79 $\pm$ 0.06  & 14.83 $\pm$ 0.04  & 13.77 $\pm$ 0.05  & 17.26 $\pm$ 0.12  & 15.96 $\pm$ 0.09 & 17.21 $\pm$ 0.12  \\
2016-06-07  & 546.6     & $-$1.6         &14.73 $\pm$ 0.05  & 14.79 $\pm$ 0.04  & 13.77 $\pm$ 0.05  & 17.46 $\pm$ 0.13  & 16.08 $\pm$ 0.08 & 16.91 $\pm$ 0.10  \\
2016-06-08  & 547.9     & $-$0.3         &14.84 $\pm$ 0.05  & 14.80 $\pm$ 0.04  & 13.68 $\pm$ 0.04  & 17.43 $\pm$ 0.16  & 16.17 $\pm$ 0.08 & 17.08 $\pm$ 0.10  \\
2016-06-10  & 549.8     & $+$1.6         &14.97 $\pm$ 0.05  & 14.81 $\pm$ 0.04  & 13.66 $\pm$ 0.04  & 17.85 $\pm$ 0.12  & 16.28 $\pm$ 0.07 & 17.23 $\pm$ 0.09  \\
2016-06-11  & 551.0     & $+$2.8         &15.05 $\pm$ 0.05  & 14.82 $\pm$ 0.04  & 13.63 $\pm$ 0.04  & 17.78 $\pm$ 0.12  & 16.36 $\pm$ 0.08 & 17.29 $\pm$ 0.10  \\
2016-06-12  & 552.3     & $+$4.2         &15.08 $\pm$ 0.06  & 14.84 $\pm$ 0.04  & 13.64 $\pm$ 0.04  & 18.15 $\pm$ 0.20  & 16.26 $\pm$ 0.08 & 17.51 $\pm$ 0.13  \\
2016-06-12  & 552.4     & $+$4.2         &15.10 $\pm$ 0.07  & 14.89 $\pm$ 0.05  & 13.67 $\pm$ 0.05  & 17.92 $\pm$ 0.18  & 16.40 $\pm$ 0.11 & 17.65 $\pm$ 0.16  \\
2016-06-14  & 554.4     & $+$6.2         &15.41 $\pm$ 0.05  & 15.01 $\pm$ 0.04  & 13.71 $\pm$ 0.04  & 18.06 $\pm$ 0.14  & 16.66 $\pm$ 0.08 & 17.58 $\pm$ 0.19  \\
2016-06-15  & 554.9     & $+$6.8         &15.48 $\pm$ 0.06  & 15.02 $\pm$ 0.04  & 13.66 $\pm$ 0.04  & 18.03 $\pm$ 0.15  & 16.68 $\pm$ 0.09 & 17.55 $\pm$ 0.11  \\
2016-06-16  & 556.0     & $+$7.8         &15.49 $\pm$ 0.06  & 15.02 $\pm$ 0.04  & 13.77 $\pm$ 0.04  & 18.30 $\pm$ 0.20  & 16.78 $\pm$ 0.10 & 17.53 $\pm$ 0.12  \\
2016-06-18  & 558.1     & $+$9.9         &15.75 $\pm$ 0.07  & 15.27 $\pm$ 0.04  & 13.81 $\pm$ 0.04  & 17.99 $\pm$ 0.15  & 17.07 $\pm$ 0.12 & 18.09 $\pm$ 0.18  \\
2016-06-21  & 561.0     & $+$12.9        &15.49 $\pm$ 0.04  & 15.49 $\pm$ 0.04  & 14.01 $\pm$ 0.04  & 18.37 $\pm$ 0.16  & 17.12 $\pm$ 0.10 & 18.20 $\pm$ 0.16  \\
\hline
\end{tabular}\\
$^{\ast}$ With reference to the $B$-band maximum (JD~2457548.2).
\end{table*}


\begin{table}
\centering
\caption{Log of spectroscopic observations of ASASSN-16fp.} 
\begin{tabular}{cccc} \hline
   Date      &     J.D.   & Phase$^{\ast}$& Range       \\        
(yyyy-mm-dd) & (2457000+) & (d)           & (\AA)       \\        
\hline
2016-05-30   & 539.4      & $-$8.8        &  3500--9250 \\        
2016-05-31   & 540.4      & $-$7.8        &  3500--9250 \\        
2016-06-01   & 541.4      & $-$6.8        &  3500--9250 \\        
2016-06-02   & 542.4      & $-$5.8        &  3500--7800 \\        
2016-06-03   & 543.4      & $-$4.8        &  3500--9250 \\        
2016-06-10   & 550.3      & $+$2.2        &  3500--9250 \\        
2016-06-11   & 551.4      & $+$3.2        &  3500--9250 \\        
2016-06-13   & 553.4      & $+$5.2        &  3500--9250 \\        
2016-06-17   & 557.4      & $+$9.3        &  3500--9250 \\        
2016-06-19   & 559.3      & $+$11.1       &  3500--9250 \\        
2016-06-21   & 561.4      & $+$13.2       &  3500--9250 \\        
2016-06-24   & 564.4      & $+$16.3       &  3500--9250 \\        
2016-06-26   & 566.3      & $+$18.1       &  3500--9250 \\        
2016-06-30   & 570.4      & $+$22.2       &  3500--9250 \\        
2016-07-01   & 571.4      & $+$23.3       &  3500--9250 \\        
2016-07-06   & 576.2      & $+$28.0       &  3500--9250 \\        
2016-07-11   & 581.4      & $+$33.3       &  5200--9250 \\        
\hline
\end{tabular} \\
$^{\ast}$ With reference to the $B$-band maximum (JD 2457548.2). 
\label{tab_spec}
\end{table}

\begin{figure}
\centering
\includegraphics[scale = 0.145]{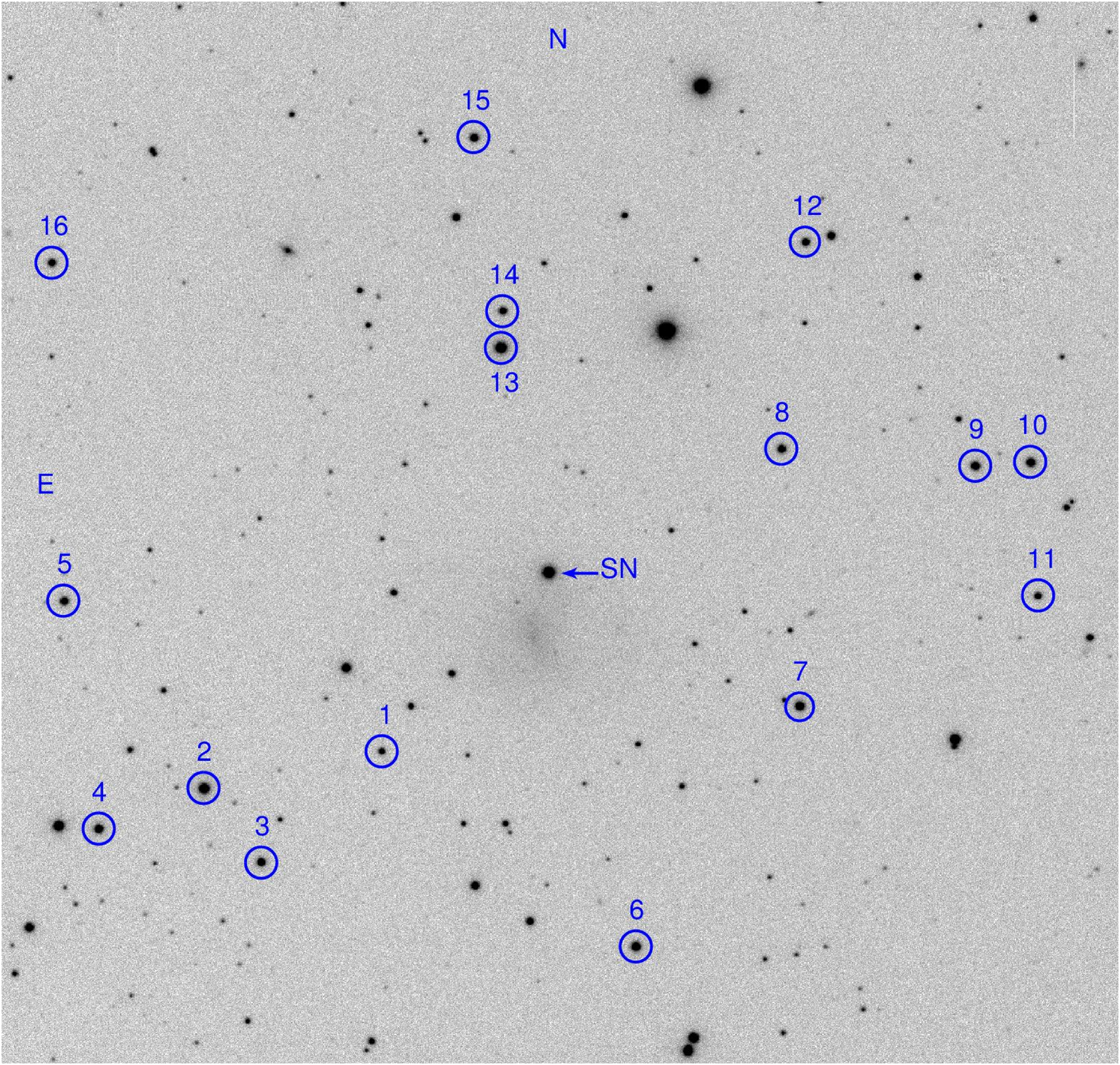}
\caption{Identification chart of ASASSN-16fp. The $R$-band image obtained using the HCT on 2016 June 13 is shown.
The field of view is roughly 9.5 arcmin $\times$ 9.5 arcmin. The SN is marked and local secondary stars are
numbered with IDs 1 to 16 (see Table~\ref{tab_id} for their magnitudes). North is up and East is to the left.}
\label{fig_1}
\end{figure}

\subsection{Photometric observation}\label{obs_ph}

Optical monitoring of ASASSN-16fp began on 2016 May 29 (JD~2457538.4). The observations were carried out 
in Bessell $UBVRI$ bands with the Himalayan Faint Object Spectrograph Camera (HFOSC) mounted on the $f/9$ Cassegrain 
focus of the 2 m Himalayan Chandra Telescope (HCT) situated at the Indian Astronomical Observatory (IAO), 
Hanle, India.
HFOSC is equipped with liquid nitrogen cooled 2k $\times$ 4k pixels SITe CCD chip (pixel size 15 $\times$ 15 $\mu$m). 
The gain and readout noise of the detector are 1.22 e$^{-}$/ADU and 4.87e$^{-}$, respectively. 
With a plate scale of 0.296 arcsec per pixel, the central 2k $\times$ 2k region covers a field of 
10$\arcmin \times$ 10$\arcmin$ on the sky and the same was used for imaging purpose. We obtained several bias 
and twilight flat frames in addition to the science frames. The pre-processing of raw data (bias-subtraction, 
flat-fielding and cosmic ray removal) was performed using the standard tasks available in the data reduction 
software {\small IRAF\,}\footnote{{\small IRAF} is distributed by the National Optical Astronomy 
Observatory, which is operated by the Association of Universities for Research in Astronomy (AURA) under a cooperative 
agreement with the National Science Foundation.}. In order to increase the signal-to-noise ratio, on some of the nights,
multiple frames were taken and co-added in respective bands after alignment of the images.


To calibrate a sequence of secondary standards in the SN field, we observed Landolt photometric standard 
fields \citep{1992AJ....104..340L} PG~1633+009, PG~2213-006 and SA~110 on four nights under good photometric 
conditions along with the SN field. The point spread function photometry on all frames (including the SN and Landolt 
fields) was done using the stand-alone version of {\small DAOPHOT}\footnote{{\small DAOPHOT} stands for Dominion 
Astrophysical Observatory Photometry.} \citep{1987PASP...99..191S, 1992ASPC...25..297S}.
The average atmospheric extinction values in $U$, $B$, $V$, $R$ and $I$ bands for the site were adopted from 
\citet{2008BASI...36..111S}. 
The observed Landolt field stars having a brightness range of 12.02 $\le V \le$ 16.25 mag and colour range of 
$-0.22 \le B-V \le 2.53$ mag were chosen for the calibration. Using these stars, transformation to the 
standard system was derived by applying the average colour terms and the photometric zero points. 
Sixteen secondary standard stars calibrated in this way are marked in Figure~\ref{fig_1} and their respective $UBVRI$
magnitudes averaged over four nights are listed in Table~\ref{tab_id}. 
Since, the location of SN in its host galaxy is fairly isolated, we did not consider host galaxy flux contamination
to the SN. Final results of the SN photometry in $UBVRI$ bands are listed in Table~\ref{tab_hct}.

The {\it Swift UVOT} data on ASASSN-16fp, downloaded from the {\it Swift} database 
{(\url{http://www.swift.ac.uk/swift_portal/})} are also used in this work. 
The SN was monitored in the UVOT $uvw2$, $uvm2$, $uvw1$, $u$, $b$, and $v$ bands 
\citep[see,][]{2008MNRAS.383..627P}. 
These observations began on JD 2457536.2 and continued  up to JD 2457581.4.
The prescriptions of \citet{2009AJ....137.4517B} were adopted to reduce the UVOT data.
Aperture photometry was performed to estimate the SN magnitudes, using {\it uvotsource} 
task in HEASoft (High Energy Astrophysics Software).
It is recommended using an aperture size of 5$\arcsec$ to estimate the magnitudes.
However, since the SN was faint, we used a smaller aperture size of 3$\arcsec$, and applied 
aperture corrections as listed by \citet{2008MNRAS.383..627P}.
The background counts were determined from the nearby region having an aperture size similar to that used for
the SN. Final {\it Swift} magnitudes (in Vega-system) are listed in Table~\ref{tab_swift}.


\subsection{Spectroscopic observation}\label{obs_sp}

Low-resolution optical spectroscopic observations of ASASSN-16fp were obtained at seventeen epochs during 
2016 May 30 (JD~2457539.4) to 2016 July 11 (JD~2457581.4). All spectra were obtained using grisms Gr\#7 (3500--7800 \AA) 
and Gr\#8 (5200--9250 \AA) with the HFOSC, having a resolution of $\sim$7 \AA. The journal of spectroscopic 
observations is given in Table~\ref{tab_spec}. 
Calibration frames such as spectra of arc lamp and spectrophotometric standards were also obtained. Spectroscopic data 
reduction was done under the {\small IRAF} environment. Bias and flat field corrections were performed on each frame, 
and the one dimensional spectra extracted using the optimal extraction method.
The dispersion solutions obtained using arc lamp spectra were used for wavelength calibration.
To secure the wavelength calibration, night-sky emission lines were used, and wherever necessary small shifts were
applied. 
Instrumental response curves, derived using the spectrophotometric standards observed on most of the nights,
were used for flux calibration. For those nights where standard star observations were not available,
the response curves obtained during nearby nights were used. To construct a single flux calibrated spectrum, the 
flux calibrated spectrum in each grism was combined. The spectra were then scaled with respect to the calibrated $UBVRI$ 
fluxes to bring them to an absolute flux scale. Finally, the SN spectra were corrected for the host galaxy redshift 
of $z$ = 0.0037 (from NED), and de-reddened by a total reddening $E(B-V)$ = 0.074 mag (see Section~\ref{color}).

\begin{table*}
\centering
\caption{Estimated light curve parameters of ASASSN-16fp.}
\label{tab_p}
\small
\begin{tabular}{lccccc}
\hline
Parameter               &         $U$       &        $B$        &        $V$        &        $R$        &         $I$       \\ \hline
JD max (2457000+)       &  545.8 $\pm$ 0.5  &  548.2 $\pm$ 0.4  &  550.7 $\pm$ 0.4  &  553.2 $\pm$ 0.4  &  555.3 $\pm$ 0.5  \\
T$_{peak}^{\#}$ (days)  &  --3.4 $\pm$ 0.6  &        0          &  +2.5  $\pm$ 0.6  &  +5.0  $\pm$ 0.6  &  +7.1  $\pm$ 0.6  \\
Apparent mag at max     &  14.84 $\pm$ 0.02 &  14.76 $\pm$ 0.01 &  13.74 $\pm$ 0.01 &  13.50 $\pm$ 0.01 &  13.54 $\pm$ 0.01 \\
$\Delta$m$_{15}$        &  1.06  $\pm$ 0.04 &  1.01  $\pm$ 0.03 &  0.60  $\pm$ 0.02 &  0.42  $\pm$ 0.02 &   0.32 $\pm$ 0.02 \\
\hline
\end{tabular}\\
$^{\#}$ Time to reach the maximum for each band with respect to the $B$-band maximum.
\end{table*}

\section{Light curve properties}\label{lc}

A discussion on the epoch of explosion, the multi-band light curves, colour evolution and absolute $V$-band light curve 
of ASASSN-16fp is provided in this section. These properties are compared with other well studied Ic and BL-Ic events.

\subsection{Explosion epoch}

Given the fact that we have good cadence of data points during the pre-maximum phase, we estimated the 
explosion epoch by fitting a power law of the form $L(t) = C\, \times\, (t - t_{0})^{\alpha}$ to 
the initial data points (pre-maximum). Here, $L$ denotes the luminosity at time $t$, the parameter 
$C$ defines the rising rate, $t_{0}$ is the time of explosion and $\alpha$ is a free parameter.
The best fit to the observed data points provides an explosion date of 2016 May $26.8 \pm 0.6$ (JD~2457535.3). 
ASASSN-16fp was detected on 2016 May 27.6 (see Section~\ref{intro}), with pre-outburst last non-detection 
reported on 2016 May 21.5 \citep[see][]{2016ATel.9086....1H}. 
Assuming the SN exploded between the last non-detection and the date of discovery, the explosion date 
may be estimated as 2016 May 25 (JD~2457533.5 $\pm$ 3). 
We adopt an average of the above two i.e. 2016 May 25.9 (JD 2457534.4) as our estimate to the date 
of the explosion of ASASSN-16fp. This is consistent with the explosion epoch of JD 2457533.0 estimated 
by \citet{2017ApJ...837....1Y}.
%

\begin{table*}
\centering
\caption{Basic information of the SNe sample used in this study.}
\begin{tabular}{lcccc} \hline
   SN name     &     Type          &  $E(B-V)_{tot}$  & Distance modulus    & References     \\
               &                   &    (mag)         & (mag)               &                \\
\hline
    1997ef     &    BL-Ic          &     --           &   33.63             & 1              \\
    1998bw     &    BL-Ic (GRB-SN) &    0.05          &   32.28             & 2, 3, 4        \\
    2002ap     &    BL-Ic          &    0.08          &   29.50             & 5, 6, 7, 8, 9  \\
    2003jd     &    BL-Ic          &    0.14          &   34.46             & 10             \\
    2004aw     &    Tr-Ic          &    0.37          &   34.17             & 11             \\
    2006aj     &    BL-Ic (XRF-SN) &    0.13          &   35.60             & 12, 13, 14     \\
    2007ru     &    BL-Ic          &    0.26          &   34.15             & 15             \\
    2007gr     &    Ic             &    0.09          &   29.84             & 16, 17         \\
    2009bb     &    BL-Ic          &    0.58          &   33.01             & 18             \\
    2010bh     &    BL-Ic (XRF-SN) &    0.49          &   36.90             & 19, 20         \\
    2012ap     &    BL-Ic          &    0.45          &   33.17             & 21             \\
    2014ad     &    BL-Ic          &    0.24          &   32.11             & 22             \\
\hline
\end{tabular}
\begin{minipage}{\textwidth}
References:
$^{1}$ \citep{2000ApJ...545..407M}, $^{2}$ \citep{1998Natur.395..672I}, $^{3}$ \citep{1998Natur.395..670G},
$^{4}$ \citep{2011AJ....141..163C}, $^{5}$ \citep{2002MNRAS.332L..73G}, $^{6}$ \citep{2003PASP..115.1220F},
$^{7}$ \citep{2002ApJ...572L..61M}, $^{8}$ \citep{2003MNRAS.340..375P}, $^{9}$ \citep{2003ApJ...592..467Y},
$^{10}$ \citep{2008MNRAS.383.1485V}, $^{11}$ \citep{2006MNRAS.371.1459T}, $^{12}$ \citep{2006ApJ...643L..99M},
$^{13}$ \citep{2006Natur.442.1011P}, $^{14}$ \citep{2006A&A...454..503S}, $^{15}$ \citep{2009ApJ...697..676S},
$^{16}$ \citep{2009A&A...508..371H}, $^{17}$ \citep{2008ApJ...673L.155V}, $^{18}$ \citep{2011ApJ...728...14P},
$^{19}$ \citep{2011ApJ...740...41C}, $^{20}$ \citep{2012A&A...539A..76O}, $^{21}$ \citep{2015ApJ...799...51M},
$^{22}$ \citep{sahu_sn14ad}.
\end{minipage}
\label{sn_inf}
\end{table*}


\subsection{Light curve features}\label{lc_fit}

\begin{figure}
\centering
\includegraphics[scale = 0.432]{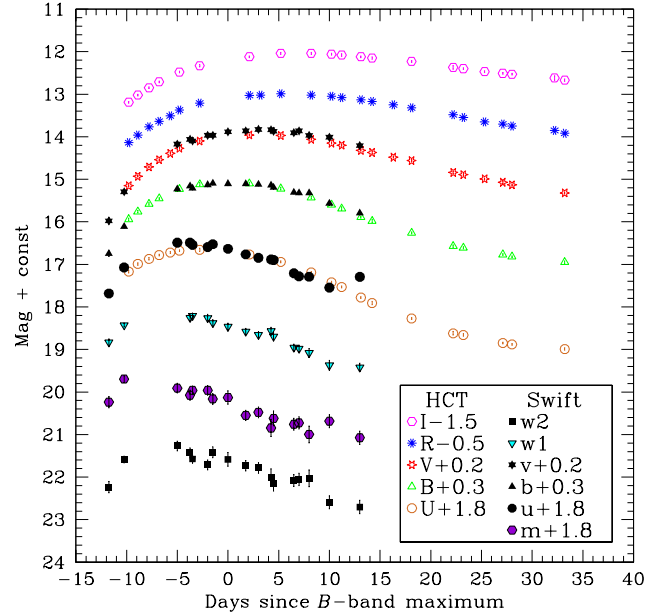}
\caption{The observed light curves of ASASSN-16fp in optical $UBVRI$ and $uvw2, uvm2, uvw1$ and $u, b, v$
NUV-optical ({\it Swift} UVOT) bands. For clarity, the light curves in different bands have been shifted 
vertically by the indicated amount.}
\label{fig_lc}
\end{figure}

The $UBVRI$ and {\it Swift} UVOT light curves of ASASSN-16fp are shown in Figure~\ref{fig_lc}.  
The follow-up observations began on 2016 May 29 (JD~2457638.4) and continued until around 30 d 
after $B$-band maximum. 
During this time span, we obtained photometric observations at 24 epochs with the HCT, and at
17 epochs with {\it Swift} satellite.


\begin{figure}
\centering
\includegraphics[scale = 0.43]{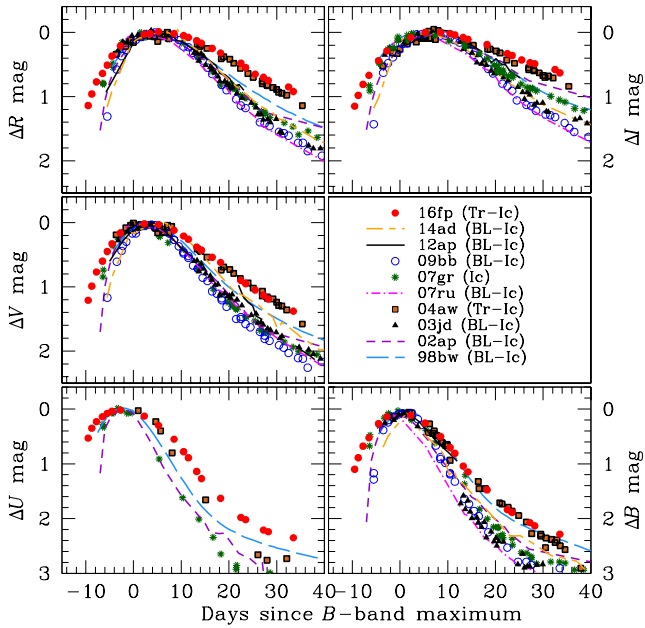}
\caption{Comparison of $UBVRI$ light curves of ASASSN-16fp (indicated with red dot symbols) with other well
observed BL-Ic events from the literature SN~2014ad, 2012ap, 2009bb, 2007ru, 2003jd, 2002ap and 1998bw along
with transitional Ic SN~2004aw and normal Ic SN~2007gr. The compared light curves of the SNe have been shifted 
arbitrarily to match the date of maximum and magnitude at maximum in the respective bands.}
\label{fig_lccomp}
\end{figure}

The epoch of maximum magnitude in different bands and their corresponding brightness were determined by 
fitting a low-order polynomial to the observed data points around the maximum light. The derived values are 
listed in Table~\ref{tab_p}. The light curve evolution of ASASSN-16fp follows the behaviour exhibited by other 
Ic/BL-Ic SNe i.e. the peak in the bluer bands appear earlier than the redder bands, which is attributed to  
the underlying spectrum and/or cooling of the photosphere \citep{2015A&A...574A..60T}.
The maximum in the $B$-band occurred on JD 2457548.2 $\pm$ 0.4 (i.e. around 14 d after the explosion) 
with an apparent magnitude of 14.76 $\pm$ 0.01 mag while, in the $U$-band, it peaked 3.4 days earlier. It 
took around 2.5, 5 and 7 days after $B$-band maximum to reach to maximum light in the $V$, $R$ and $I$ bands, 
respectively (cf. Table~\ref{tab_p}).

In Figure~\ref{fig_lccomp}, the $UBVRI$ light curves of ASASSN-16fp are plotted along with those of other well 
studied Type Ic and BL-Ic SNe (see Table~\ref{sn_inf}). The light curves of SNe used for the comparison were 
shifted in time axis to match the epoch of $B$-band maxima of ASASSN-16fp, and normalized with respect to their 
peak magnitudes. For the purpose of comparison, only those objects that have densely sampled light curves, with 
coverage around maximum light were considered.
The $U$-band light curve of ASASSN-16fp is compared with those of SN~2002ap, SN~2004aw, SN~2007gr and SN~1998bw, 
as only these objects fit in our criteria.

A careful inspection of Figure~\ref{fig_lccomp} reveals that the light curves of ASASSN-16fp is broad in all the bands.
Both the rise and decline of the light curves are slower than most of the objects used in comparison. 
The evolution of $U$-band light curve of ASASSN-16fp between 0 to +20 d is slower than SN~1998bw, 2002ap, 2004aw and 2007gr.
The post-maximum evolution of $BVRI$ band light curves of SN~2004aw and ASASSN-16fp are similar.
The $\Delta$m$_{15}$ (decline in magnitude after 15 days post-maximum) parameter of ASASSN-16fp and 
SN~2004aw are comparable, except for $U$-band.
The respective $\Delta$m$_{15}$ values in $U$, $B$, $V$, $R$ and $I$ bands for ASASSN-16fp are 1.06 $\pm$ 0.04, 
1.01 $\pm$ 0.03, 0.60 $\pm$ 0.02, 0.42 $\pm$ 0.02 and 0.32 $\pm$ 0.02 and for SN~2004aw 
1.62 $\pm$ 0.25, 1.09 $\pm$ 0.04, 0.62 $\pm$ 0.03, 0.41 $\pm$ 0.03 and 0.34 $\pm$ 0.03 \citep{2006MNRAS.371.1459T}.
The $\Delta$m$_{15,R}$ of ASASSN-16fp is slightly lower than the mean $\Delta$m$_{15,R}$ 
computed by \citet{2011ApJ...741...97D} for a sample of BL-Ic (0.6 $\pm$ 0.14 mag) and normal Ic (0.73 $\pm$ 0.27 
mag) events. This emphasizes the slow decline of ASASSN-16fp light curves compared to normal Ic and BL-Ic events, while
being similar to the transitional Ic SN 2004aw.

\begin{figure}
\centering
\includegraphics[scale = 0.42]{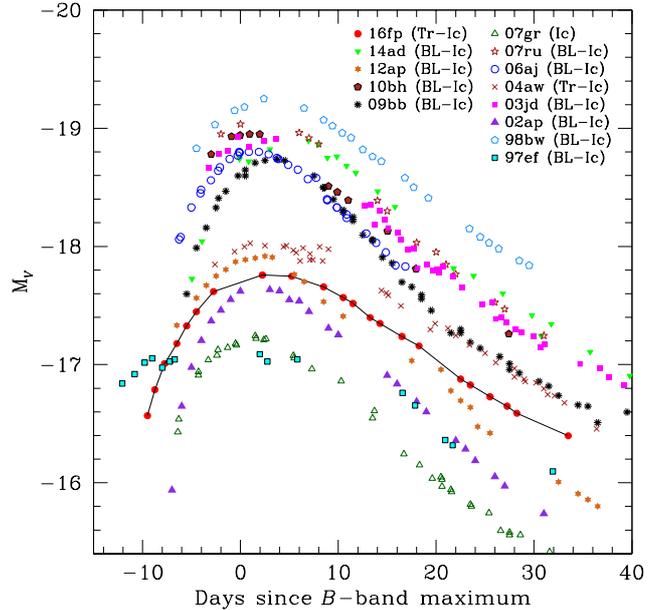}
\caption{Comparison of the $M_V$ light curve of ASASSN-16fp (shown with red dot symbols and connected with solid line) 
with other broad line Ic events: SN~2014ad, 2012ap, 2010bh, 2009bb, 2007ru, 2007gr, 2006aj, 2004aw, 2003jd, 2002ap, 
1998bw and 1997ef.}
\label{fig_4}
\end{figure}

In Figure~\ref{fig_4}, we compare the peak absolute magnitudes in $V$-band ($M_V$) for various well studied Ic 
and BL-Ic SNe. The $M_V$ is estimated after correcting for extinction and adopting distance measurements from the 
respective studies as listed in Table~\ref{sn_inf}.  
In case of the host galaxy UGC~11868 of ASASSN-16fp, precise distance measurement (e.g. cepheid distances) is 
unavailable. However, for the present study, we have adopted distance\footnote{The distance was taken from NASA/IPAC 
Extragalactic Data base \url{http://ned.ipac.caltech.edu/}, and it is corrected for Virgo infall.}
of 18.1\,$\pm$\,1.3 Mpc ($\mu$ = 31.29\,$\pm$\,0.15 mag, H$_{0}$ = 73 km sec$^{-1}$ Mpc$^{-1}$). 
With $M_V$ = $-17.70 \pm 0.2$ mag, the absolute $V$-band luminosity of ASASSN-16fp is comparable with BL-Ic
SN~2012ap \citep[$M_V = -17.9\, \pm\, 0.10$ mag,][]{2015ApJ...799...51M}, SN~2002ap \citep[$M_V = -17.37 \pm 0.05$ mag,][]
{2003PASP..115.1220F,2003MNRAS.340..375P} and transitional Ic SN~2004aw \citep[$M_V = 18.02 \pm 0.39$ 
mag,][]{2006MNRAS.371.1459T}.
From figure~\ref{fig_4} it is evident that there are several BL-Ic events considerably brighter than ASASSN-16fp, 
while BL-Ic SN~1997ef \citep[$M_{V} = -17.14$ mag,][]{2004ApJ...614..858M} and normal Ic SN~2007gr \citep[$M_{V} = 
-17.22 \pm 0.18$ mag,][]{2009A&A...508..371H} are $\sim$0.5 mag fainter than ASASSN-16fp. 
The absolute $R$-band luminosity of $M_R$ = $-17.96 \pm 0.15$ mag for ASASSN-16fp is $\sim$1 mag fainter
than BL-Ic sample ($M_{R} = -19.0$ mag) of \citet{2011ApJ...741...97D}, but only slightly fainter than
their normal Ic sample ($M_{R} = -18.3$ mag).

\subsection{Extinction and colour evolution}\label{color}

The Galactic reddening towards ASASSN-16fp is $E(B-V)$ = 0.074 $\pm$ 0.002 mag \citep*{2011ApJ...737..103S}.
The strength of Na\,{\sc i} D absorption lines is found to be coupled with the reddening \citep{1990A&A...237...79B,
2003fthp.conf..200T}, though exceptions exist \citep{2011MNRAS.415L..81P}.
In the low-resolution spectra presented in this work, a weak Na\,{\sc i} D absorption 
due to the ISM in the Milky Way is seen, while none detected at the redshift of the host galaxy.
The absence of Na\,{\sc i} D line at rest wavelength of the host indicates negligible 
extinction within the host galaxy. A similar conclusion has also been arrived at by 
\citet{2017ApJ...837....1Y}. We have used $E(B-V)$ = 0.074 mag for further analysis in this paper.

\begin{figure}
\centering
\includegraphics[scale = 0.42]{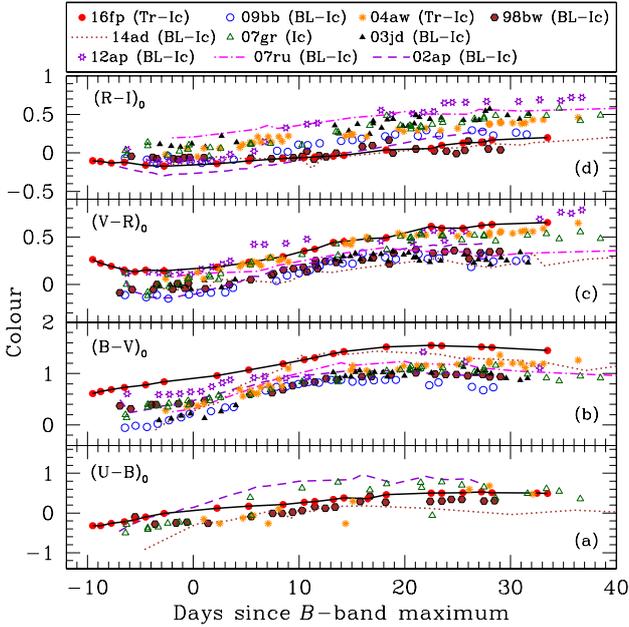}
\caption{The evolution of $(U-B)_0$, $(B-V)_0$, $(V-R)_0$ and $(R-I)_0$ colours of ASASSN-16fp.
For comparison, the similar colour curves of SN~2014ad, SN~2012ap, SN~2009bb, SN~2007ru, SN~2006aj, SN~2004aw,
SN~2003jd, SN~2002ap and SN~1998bw are also over-plotted. The ASASSN-16fp colours are shown with red dot symbols
and connected with solid line. The bibliographic sources are the same as mentioned in the text (Section~\ref{lc}).}
\label{fig_3}
\end{figure}

Figure~\ref{fig_3} presents the evolution of the reddening corrected $U-B$, $B-V$, $V-R$ and $R-I$ colours
for SNe ASASSN-16fp, 2014ad, 2012ap, 2009bb, 2007gr, 2007ru, 2004aw, 2003jd, 2002ap and 1998bw.
The extinction values for these events have been taken from the references as mentioned in Table~\ref{sn_inf}.
The diverse nature of colour evolution is likely due to the varying nature of the expansion and cooling of 
the SN photosphere.
Both $(U-B)_0$ and $(B-V)_0$ colours of majority of the SNe follow a blue to red trend until about 
$\sim$+10 days, and beyond that, the colour evolution is almost flat 
(cf. panel a and b in Fig.~\ref{fig_3}). It is to be noted that among the compared events, 
the $(B-V)_0$ colour of ASASSN-16fp is redder by $\sim$0.6 mag and, continues to be so till the last 
epoch of data presented here.  
The $(V-R)_0$ and $(R-I)_0$ colours are shown in panels (c) and (d) of Figure~\ref{fig_3}, respectively.
The $(V-R)_0$ color of ASASSN-16fp exhibits red-blue-red transition between $-10$ d to $+10$ d, a 
colour variation not evident in the other events used for comparison.  
\citet{2011ApJ...741...97D} have examined the multi-band light curve of Ibc SNe in the local universe
(distance $\sim$150 Mpc). They have shown that the $(V-R)_0$ colour of Ibc SNe around 10 days past 
$V$-band maximum shows very small scatter, ranging between 0.18 to 0.34 mag (with a mean $(V-R)_0$ 
colour 0.26 $\pm$ 0.06 mag).
Using a set of theoretical models \citet{2016MNRAS.458.1618D} have confirmed the findings of 
\citet{2011ApJ...741...97D}. At similar epoch, the $(V-R)_0$ colour of ASASSN-16fp is $\sim$0.43 
$\pm$ 0.02 mag indicating it to be intrinsically redder.
The $(R-I)_0$ colour evolution of compared SNe shows considerable dispersion. 

\begin{figure}
\centering
\includegraphics[scale = 0.421]{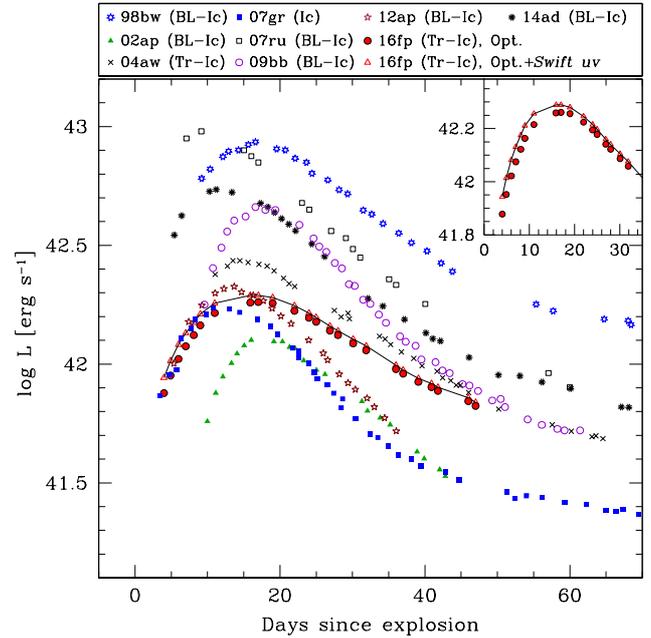}
\caption{The quasi-bolometric light curve of ASASSN-16fp compared with other Type Ic and BL-Ic events. 
Optical and optical + {\it Swift} UV bolometric light curves of ASASSN-16fp are over-plotted with
different symbols. For clarity, optical + {\it Swift} UV light curve of ASASSN-16fp is connected with solid line 
and also shown in inset.}
\label{bol_mag}
\end{figure}

\subsection{Bolometric light curve and explosion parameters}\label{bol}

The bolometric light curve is crucial for calculation of various explosion parameters such as amount of 
ejecta mass expelled during the explosion ($M_{\rm ej}$), kinetic energy ($E_{\rm k}$), mass of $^{56}$Ni 
synthesized in the explosion. These in turn help to infer the nature of the progenitor.
The quasi-bolometric light curve of ASASSN-16fp was constructed using the optical $(UBVRI)$ and {\it Swift} 
UVOT magnitudes presented in Section~\ref{lc}. The data were interpolated wherever it was necessary and
observed magnitudes were corrected for extinction $E(B-V)$ = 0.074 mag. The extinction corrected magnitudes 
were then converted to monochromatic flux at the filter effective wavelength, using the magnitude to 
flux conversion zero points from \citet{1998A&A...333..231B} and \citet{2008MNRAS.383..627P}. 
The fluxes were then interpolated using the spline fitting. 
To obtain the quasi-bolometric luminosity, fluxes from 0.31 $\mu$m to 0.93 $\mu$m were integrated. 
Additionally, when {\it Swift} data was available (between JD~2457536.2 to JD~2457561.0) flux was integrated 
from 0.16 $\mu$m to 0.93 $\mu$m. The quasi-bolometric luminosities estimated from both optical and optical + 
{\it Swift} UV magnitudes are shown with different symbols in Figure~\ref{bol_mag}.
The bolometric light curves of several other well studied Ic and BL-Ic SNe, estimated in a similar way 
have also been plotted in the same figure.  
The bolometric light curve of ASASSN-16fp peaked at JD~2457559.5 with log(L$_{bol}$) = 42.26 erg s$^{-1}$. 
It is fainter than GRB associated SN~1998bw, BL-Ic SNe 2014ad, 2009bb, 2007ru, and transitional Ic SN~2004aw. 
It is also brighter than BL-Ic SN~2002ap, and comparable to BL-Ic SN~2012ap and normal Ic SN~2007gr.
During the post-peak phase, the evolution of bolometric light curve of ASASSN-16fp is slower than SN~2002ap, 
2007gr and 2012ap.   


In SE-SNe the bolometric luminosity peak is mainly dominated by the decay of $^{56}$Ni and the shape is determined
by kinetic energy ($E_{\rm k}$) and total mass of the ejecta ($M_{\rm ej}$). To derive explosion parameters for 
ASASSN-16fp we followed the analytical model originally proposed by \citet{1982ApJ...253..785A} and 
later updated by \citet{2008MNRAS.383.1485V}. 
This model is valid only during the photospheric phase where ejecta is optically thick. 
The model involve various assumptions such as spherically symmetric and homologous expansion of the 
ejecta, centrally located and unmixed $^{56}$Ni,
small stellar radius (at the onset of explosion), and a constant opacity ($\kappa_{\rm opt}$). 

The model comprises two free parameters $M_{\rm Ni}$ and $\tau_{\rm m}$. 
The timescale ($\tau_{\rm m}$) of the light curve is represented by 

\begin{equation}\label{eq_1}
\tau_{\rm m} = \left( \frac{\kappa_{\rm opt}}{\beta c} \right)^{1/2} \left( \frac{6M^3_{\rm ej}}{5E_{\rm k}} \right)^{1/4}
\end{equation}

In the equation, $\beta \approx 13.8$ is a constant of integration \citep{1982ApJ...253..785A} and {\it c} 
is speed of light.
The optical opacity $\kappa_{\rm opt}$ is  adopted as 0.07 cm$^2$ g$^{-1}$ \citep[eg.][]{2000AstL...26..797C,
2016ApJ...818...79T}. For a uniform density \citep{1982ApJ...253..785A,1996snih.book.....A}, the kinetic energy
of ejecta can be expressed as 

\begin{equation}\label{eq_2}
E_{\rm k} \approx \frac{3}{5}\frac{M_{\rm ej}v^2_{\rm ph}}{2}
\end{equation}

\begin{figure}
\centering
\includegraphics[scale =0.115]{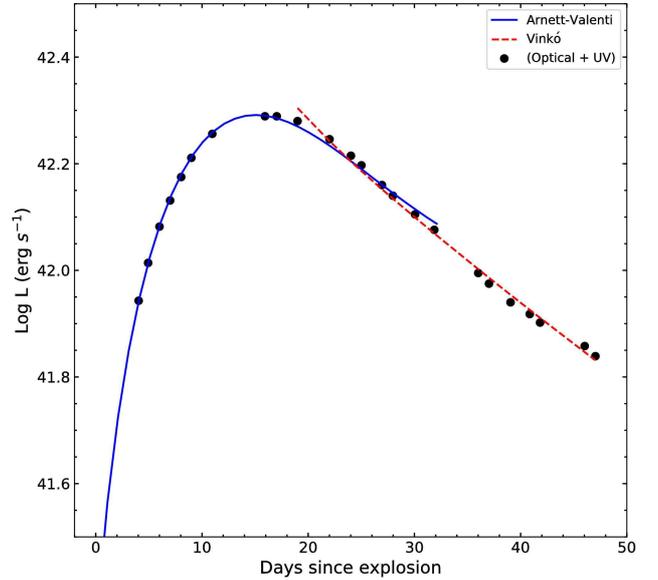}
\caption{The analytical model fitting to the bolometric light curve (optical + UV) of ASASSN-16fp (see Section~\ref{bol}).
The solid (blue) and dashed (red) lines respectively, correspond to the best fitted Arnett-Valenti and Vink{\'o} models.} 
\label{bol_fit}
\end{figure}

The bolometric light curve (optical + {\it Swift} UV) till 35 days after explosion was used for fitting the 
analytical model. The best fit values obtained using model fit are $M_{\rm Ni}$ = 0.10 $\pm$ 0.01 M$_{\odot}$ and 
$\tau_{\rm m}$ = 15.9 $\pm$ 0.1 days.
The photospheric velocity around the bolometric maximum is estimated as $v_{\rm ph}$ = 16000 $\pm$ 1000 km 
s$^{-1}$ (see Section~\ref{pre_max_spc}). Using equations (\ref{eq_1}) and (\ref{eq_2}), the values of 
$M_{\rm ej}$ and $E_{\rm k}$ are estimated as 4.50 $\pm$ 0.33 M$_{\odot}$ and 6.87$^{+1.46}_{-1.27}$ $\times$ 
10$^{51}$ erg, respectively. The best-fit model is shown in Figure~\ref{bol_fit}.

In order to verify the consistency of the derived parameters, the observed bolometric light 
curve (optical + {\it Swift} UV) is also fit with the analytical model proposed by \citet{2004A&A...427..453V}. 
Similar to the Arnett-Valenti model, this model has also several assumptions e.g. homologous expansion of the ejecta,
spherical symmetric explosion and constant opacity etc.
This model is applicable only during the post-maximum phase.     
We used their model C which assumes a core-shell density structure with a constant density core
of fractional radius ($x_{0}$), and density of the surrounding shell decreasing outward as a power law with exponent $n$.
With a fixed $x_{0}$ (= 0.15), the best fit (Fig.~\ref{bol_fit}) physical explosion parameters are $M_{\rm Ni}$ = 0.12 
$\pm$ 0.01 M$_{\odot}$, $M_{\rm ej}$ = 4.3 $\pm$ 0.2 M$_{\odot}$ and $E_{\rm k}$ = 7.0 $\pm$ 1.3 $\times$ 10$^{51}$ erg.
$M_{\rm ej}$, $M_{\rm Ni}$, $\gamma$-ray opacity ($\kappa_{\gamma}$), positron opacity ($\kappa_{+}$), $v_{\rm ph}$
and $n$ were kept as free parameters for the fitting.
The explosion parameters estimated by both methods are mutually consistent.

It should be noted that no correction has been applied for the missing flux in the IR-bands. Near-infrared (NIR) flux 
contribution to the bolometric flux is found to be significant and 
varying with SN age, {\it e.g.} near the maximum brightness NIR contributes to $\sim$20 -- 25\% which increases to 
$\sim$40--50\% one month later \citep{2006ApJ...644..400T,2008MNRAS.383.1485V,2011ApJ...740...41C}. 
Including an NIR contribution of 20\% in pre-maximum, 25\% between 0 to +10 days and 31 to 45\% between +10 to +35 days 
in the UV-optical bolometric light curve of ASASSN-16fp, both analytical models were re-fitted. The explosion
parameters including the NIR are i.e. Arnett-Valenti model: $M_{\rm Ni}$ = 0.15 $\pm$ 0.01
M$_{\odot}$, $M_{\rm ej}$ = 5.8 $\pm$ 0.5 M$_{\odot}$ and $E_{\rm k}$ = 8.9 $\pm$ 1.6 $\times$ 10$^{51}$ erg and 
Vink{\'o} model: $M_{\rm Ni}$ = 0.15 $\pm$ 0.01 M$_{\odot}$, $M_{\rm ej}$ = 5.4 $\pm$ 0.2 M$_{\odot}$ and $E_{\rm k}$ = 
8.8 $\pm$ 1.9 $\times$ 10$^{51}$ erg.

\section{Spectral analysis}\label{spec_anal}

In the following sections we present the spectral evolution of ASASSN-16fp,
thanks to densely sampled spectroscopic data obtained during the photospheric phase (cf. Table~\ref{tab_spec}).
Spectral comparison with other well studied events at different epochs is also presented.
The spectroscopic data of SN~2007gr used for comparison purpose in this study have been obtained with HCT
and these are unpublished.

\begin{figure}
\centering
\includegraphics[scale = 0.45]{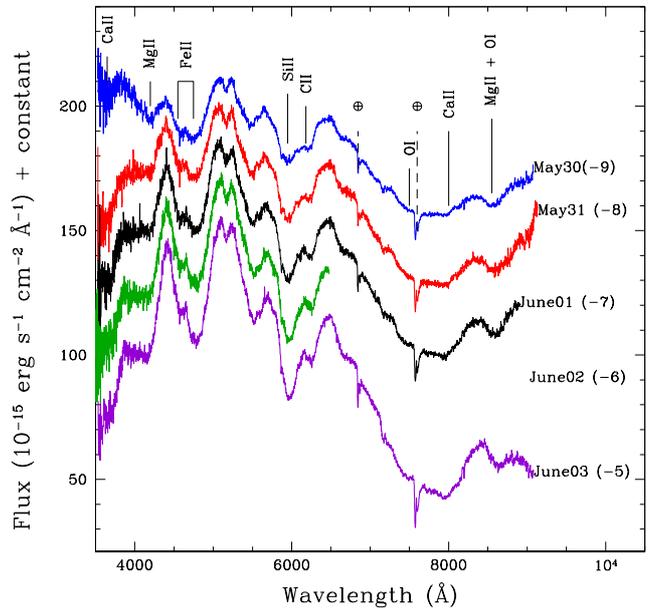}
\caption{Pre-maximum spectral evolution of ASASSN-16fp. The phases marked are respective to $B$-band maximum light.
The spectra have been corrected for redshift of the host galaxy and reddening. Vertical shifts have been applied
for clarity and prominent spectral lines are marked. The telluric lines are indicated with encircled plus symbol.}
\label{spec_pre}
\end{figure}

\begin{figure}
\centering
\includegraphics[scale = 0.42]{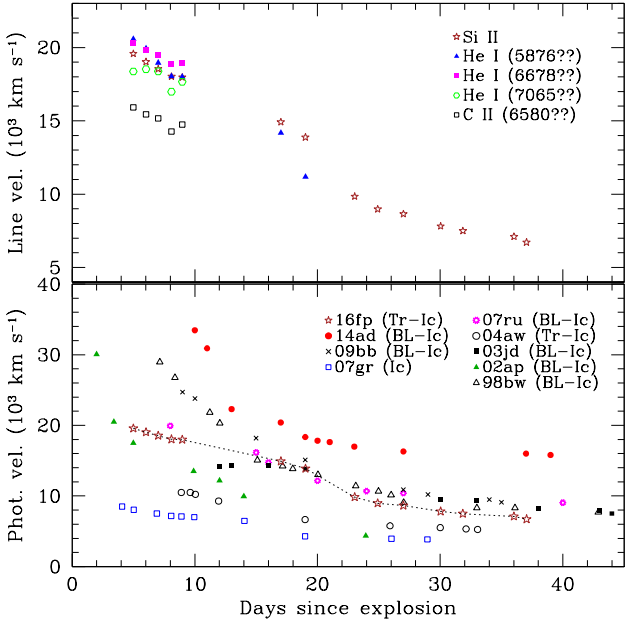}
\caption{Top panel: Evolution of the velocity of different spectral lines observed in ASASSN-16fp.
Error in these estimates was never below 800 km s$^{-1}$.
Bottom panel: The photospheric velocity of ASASSN-16fp compared with BL-Ic SNe
(2014ad, 2009bb, 2007ru, 2003jd, 2002ap and 1998bw), normal Ic SN~2007gr and transitional Ic
SN~2004aw. For better identification, the symbols of ASASSN-16fp are connected with dotted line
(see Section \ref{pre_max_spc}).}
\label{fig_c2}
\end{figure}

\subsection{Pre-maximum spectral evolution}\label{pre_max_spc}

Supernova ASASSN-16fp was monitored extensively during the pre-maximum and early post-maximum phases. The spectral 
evolution during the pre-maximum phase is presented in Figure~\ref{spec_pre}. The first spectrum obtained at 
$\sim$--9 d shows a blue continuum, with well developed features due to Ca\,{\sc ii} H \& K, Mg\,{\sc ii}, Fe\,{\sc ii},
O\,{\sc i}, Ca\,{\sc ii} NIR triplet. The features due to various species are broad, indicating high 
expansion velocity of the ejecta. The pre-maximum spectra of ASASSN-16fp also show a shallow absorption at 
$\sim$5150 \AA, and broad absorptions at $\sim$5500, 6000 and 6250 \AA\,. The shallow
absorption at $\sim$5150 \AA\, was also noticed in the spectrum of SN 2004aw obtained around maximum light 
\citep{2006MNRAS.371.1459T}, which was unidentified. The absorption at 
$\sim$6000 \AA\ can be associated with Si\,{\sc ii} 6355\AA\, \citep{2003PASP..115.1220F,2008MNRAS.383.1485V,
2009A&A...508..371H,2009ApJ...697..676S,2011ApJ...728...14P,2016ApJ...821...57D}. The absorption blue-ward of Si\,{\sc ii} 
line has been identified with Na\,{\sc i} possibly contaminated by He\,{\sc i} in other BL-Ic SNe
\citep{2003MNRAS.340..375P,2008MNRAS.383.1485V,2009A&A...508..371H,2009ApJ...697..676S}.


The absorption feature at 6250 \AA\, redward of Si\,{\sc ii} 6355\AA\ is prominently seen in the pre-maximum 
spectra obtained till $-5$ d. The next spectrum taken at +2 d shows only a small kink and beyond that the feature 
is not seen. This absorption feature was also detected in the early-phase spectrum of SN~2007gr \citep{2008ApJ...673L.155V}, 
SN~2004aw \citep{2006MNRAS.371.1459T} and SN~2013ge \citep{2016ApJ...821...57D}. 
In SN~2004aw and SN~2013ge its possible association with C\,{\sc ii} 6580 \AA\, has been indicated. In some other 
cases this line has also been identified with He\,{\sc i} 6678 \AA\, line. It is argued by \citet{2008ApJ...673L.155V}
that due to increased diffusion of $\gamma$-rays with time \citep{1998MNRAS.295..428M}, the intensity of lines due 
to He\,{\sc i} is expected to increase. The fact that this line was seen in the pre-maximum spectra of SN~2007gr
and disappeared in the spectra taken close to maximum, led them to identify this line as  C\,{\sc ii} 6580 \AA\,.
They have confirmed detection of carbon by identifying C\,{\sc ii} 7234 \AA\, in the early phase optical spectra
of SN~2007gr and strong C\,{\sc i} line (at 8335, 9094 and 9405 \AA\,) in the optical and (at 10400 \AA\,)  
NIR spectrum at around two weeks past $B$-maximum. 
As the expected position of C\,{\sc ii} 7234 \AA\, line in ASASSN-16fp is very close to the telluric absorption, 
it is difficult to identify this line.
However, we do see a broad absorption around 9000 \AA\, in the spectra of around two weeks after maximum light 
(see Section~\ref{post_pk}). It is suggested by \citet{2009A&A...508..371H} 
that the broad absorption seen $\sim$9000 \AA\, in the spectrum of SN~2004aw and SN~2003jd may be due to blending 
of the C\,{\sc i} lines (at 8335, 9094 and 9405 \AA\,). 
If the absorption redward to Si\,{\sc ii} is identified with C\,{\sc ii} 6580 \AA\, the velocity estimated using this line
is always found to be lower than the velocity measured using Si\,{\sc ii} 6355 \AA\, line (c.f. Figure~\ref{fig_c2}, top panel).

In the early phase spectra, the absorption features at 5500, 6300 and 6850 \AA\, are identified 
as He\,{\sc i} (5876, 6678 and 7065 \AA\,) by \citet{2017ApJ...837....1Y} at an expansion velocity of $\sim$18000 km 
sec$^{-1}$. Assuming this identification, the velocities estimated from our spectra are plotted in the top panel of
Figure~\ref{fig_c2}.
It is evident from the figure that the velocities estimated using these
lines match well with those inferred using the Si\,{\sc ii} 6355 \AA\, line. It is shown by \citet{2016ApJ...820...75P}
that ambiguity is always present in identifying the absorptions at $\sim$5500 \AA\, with Na\,{\sc i} and He\,{\sc i}, at
$\sim$6200 \AA\, with Si\,{\sc ii} and H\,{\sc i}, at $\sim$6300 \AA\, with C\,{\sc ii} and He\,{\sc i}.
This degeneracy can partially be removed with the help of NIR spectra. Unfortunately, we do not have
NIR spectra of ASASSN-16fp for a firm identification.

The Fe\,{\sc ii} lines provide a good estimate of the photospheric velocity, but usually, these are blended.
Therefore, Si\,{\sc ii} 6355 \AA\, line is considered as the tracer of the photospheric velocity of the ejecta 
\citep[though, contamination of some ions may persist e.g. detached He\,{\sc i} 6678 \AA, detached 
C\,{\sc ii} 6580 \AA\, and detached H$_{\alpha}$, see,][]{1996ApJ...462..462C,2006PASP..118..791B,
2006A&A...450..305E}. We used Si\,{\sc ii} line to estimate the photospheric velocity by fitting 
Gaussian profile to the absorption trough in the redshift corrected spectra.
The resulting velocities of ASASSN-16fp along with other well observed SNe are over-plotted in Figure~\ref{fig_c2} 
(bottom panel). It is worth mentioning that among BL-Ic, SN~1998bw (GRB associated), 2002ap and 2014ad show very 
high expansion velocities (reaching around 30000 km s$^{-1}$) few days after the explosion. 
In the post-maximum phase ($>$20 d), the velocity dispersion is small and a majority of the BL-Ic events stay
around $\sim$8000 km s$^{-1}$. 
The photospheric velocity of ASASSN-16fp declined from $\sim$19500 km s$^{-1}$ to $\sim$14000 km s$^{-1}$ 
during $\sim$5 d to $\sim$20 d after the explosion and beyond $\sim$25 d, it flattens around 8000 km s$^{-1}$. 
The photospheric velocities of SN~2004aw and SN~2007gr are low, and evolve at a slower rate as compared to 
BL-Ic. Among the sample presented here, the velocity of ASASSN-16fp lies between BL-Ic and Ic 
(normal and transitional) events.


\begin{figure}
\centering
\includegraphics[scale = 0.43]{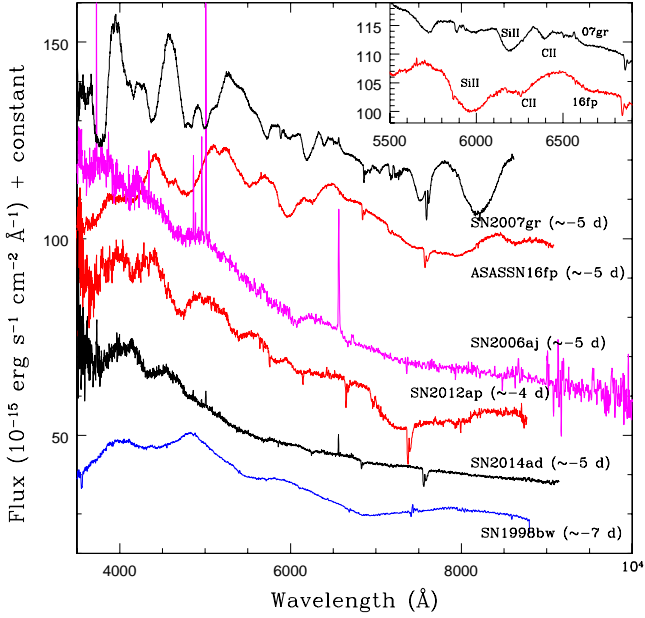}
\caption{Comparison of spectral features of ASASSN-16fp and other events taken before maximum light ($\sim$--5 days).
The spectra have been corrected for reddening and redshift. Vertical shifts have been applied for clarity.
In inset, the positions of Si {\sc ii} and C {\sc ii} lines are also shown.}
\label{comp_5}
\end{figure}

\begin{figure}
\centering
\includegraphics[scale = 0.45]{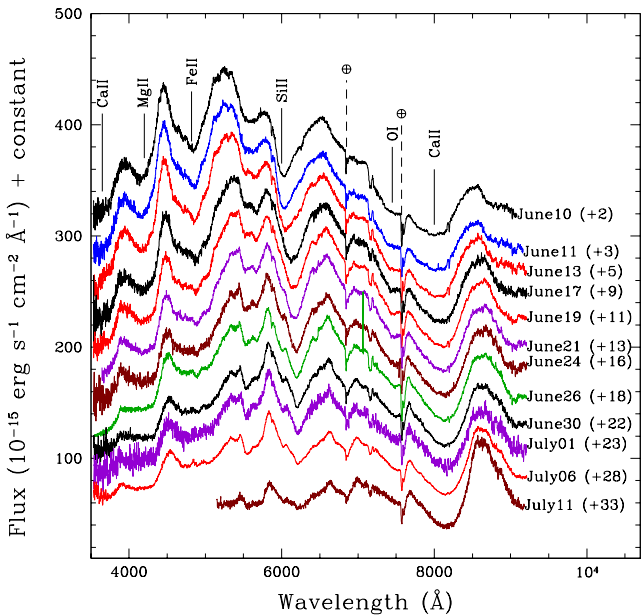}
\caption{Post-maximum phase spectral evolution of ASASSN-16fp. The host galaxy redshift and reddening corrections are
applied. Vertical shifts have been applied for clarity and prominent spectral lines are marked.  Days relative to
maximum light are indicated.}
\label{spec_post}
\end{figure}

In Figure~\ref{comp_5}, the spectrum of ASASSN-16fp at $\sim$--5 d is compared with the similar epoch spectra of other 
BL-Ic and Ic SNe. Though ASASSN-16fp has been classified as BL-Ic, the difference between the spectrum of ASASSN-16fp 
and other BL-Ic SNe is obvious. The spectra of other BL-Ic events, obtained $\sim$--5 day, show just a few broad 
absorption features, superimposed on a blue continuum. In the spectrum of ASASSN-16fp, features due to various 
species are well developed and the continuum is relatively red. The spectral features of ASASSN-16fp appears to 
be broader than the normal Type Ic SN~2007gr and narrower than other BL-Ic SNe. However, because of the high 
expansion velocity of the ejecta, features of ASASSN-16fp are more blue-shifted and the O\,{\sc i} and Ca\,{\sc ii} 
NIR triplet are heavily blended. 
The O\,{\sc i} and Ca\,{\sc ii} NIR triplet are not developed in the spectra of SN~1998bw, SN~2006aj and SN~2014ad
taken around the same epoch. The spectrum of SN~2002ap shows some signature of these lines at $\sim$--\,4 d.

\subsection{Post-maximum spectral evolution}\label{post_pk}

The post-maximum spectral evolution of ASASSN-16fp is displayed in Figure \ref{spec_post}. It includes 
spectra obtained between +2 to +33 days relative to $B$-band maximum. The spectral evolution during this phase
is characterized by the presence of continuum and absorption troughs. The appearance of such features in the 
last spectrum presented here (+33 d) implies that the SN remained in the photospheric phase until then.
Similar to the light curve, the spectral evolution is also found to be slow. Most of the features observed in 
the pre-maximum spectra continue to be present. The feature redward of Si\,{\sc ii} 6355 \AA\, line disappeared 
in the spectra obtained during the post-maximum phase. 
The Si\,{\sc ii} 6355 \AA\, absorption starts becoming narrow. The blending of O\,{\sc i} with  Ca\,{\sc ii} 
NIR triplet reduces and by $\sim$30 days, both lines appear separated, though the O\,{\sc i} line is contaminated 
by the telluric absorption. The blend of Fe\,{\sc ii} lines (4924, 5018 and 5169 \AA\,) seen as 
broad absorption at $\sim$5000 \AA\, in the pre-maximum spectra is getting resolved into its 
components. Similarly the absorption at $\sim$5500 \AA, starts splitting 
possibly into He\,{\sc i} and Na\,{\sc i}. After $\sim$+15 d, some new lines start appearing around the 
Si\,{\sc ii} 6355 \AA\, line, which have been identified by \citet{2016ApJ...820...75P} as due to O\,{\sc i}, 
Fe\,{\sc ii} and Ne\,{\sc i} \citep{2008ApJ...673L.155V}. In the red part of the spectrum, the Ca\,{\sc ii} NIR 
triplet is dominating the spectrum and progressively the emission component becomes stronger. 

\begin{figure}
\centering
\includegraphics[scale = 0.43]{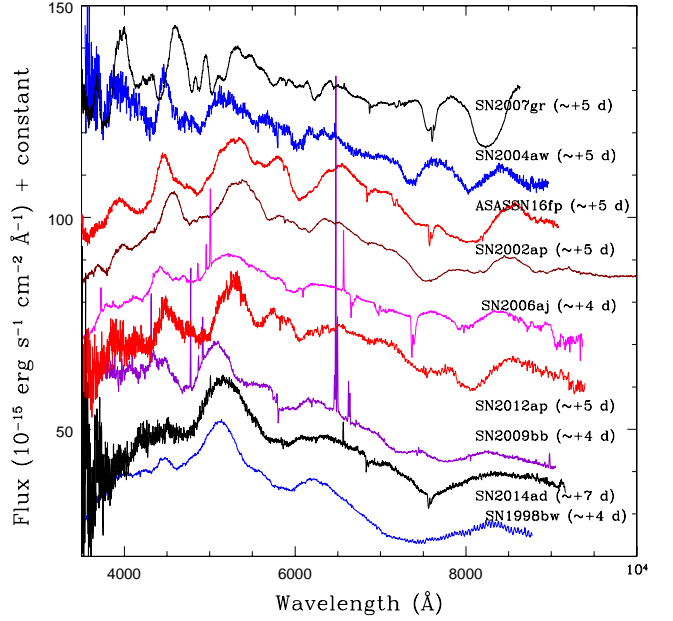}
\caption{Comparison of $\sim$+5 days spectral features of ASASSN-16fp and other events.
For clarity, the spectra have been shifted vertically.}
\label{comp+5}
\end{figure}

\begin{figure}
\centering
\includegraphics[scale = 0.43]{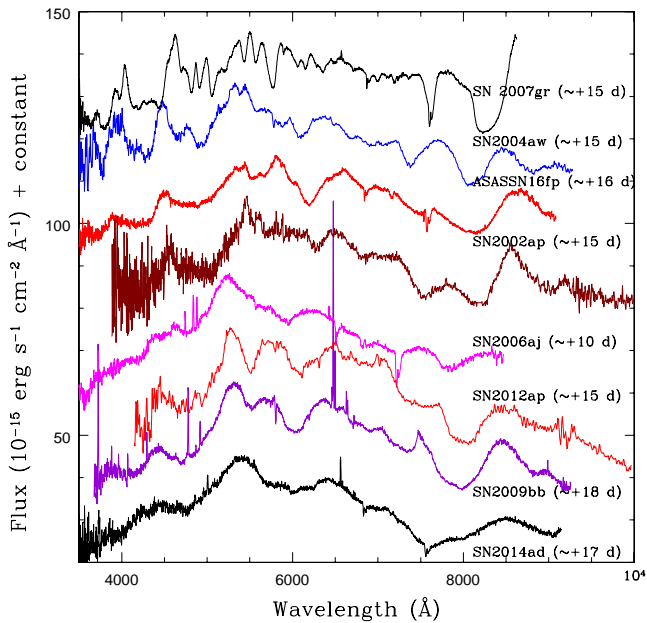}
\caption{Comparison of $\sim$+15 days spectral features of ASASSN-16fp and other events.
For clarity, the spectra have been shifted vertically.}
\label{comp+15}
\end{figure}

\begin{figure}
\centering
\includegraphics[scale = 0.43]{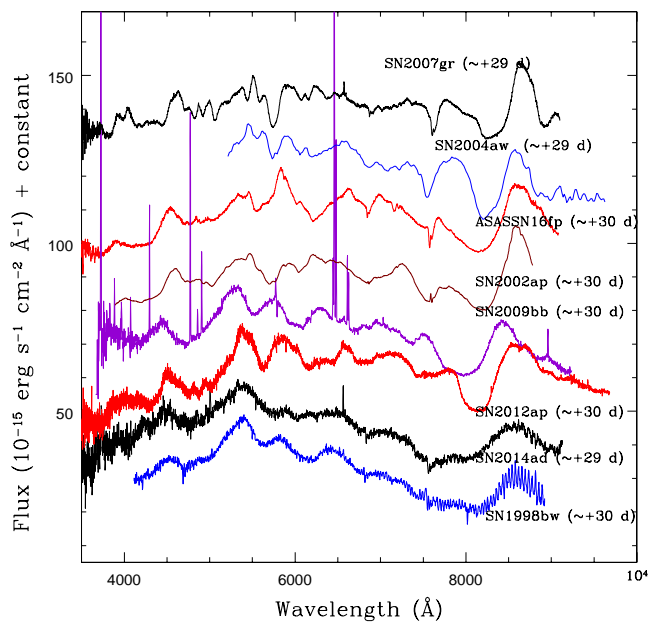}
\caption{Comparison of $\sim$+1 month spectral features of ASASSN-16fp and other events. 
For clarity, the spectra have been shifted vertically.}
\label{comp+30}
\end{figure}

In Figures \ref{comp+5}, \ref{comp+15} and \ref{comp+30} the post-maximum spectra of ASASSN-16fp obtained at 
$\sim$+5 d, $\sim$+15 d and $\sim$+30 d have been compared with other well studied Type Ic and BL-Ic SNe. 
As ASASSN-16fp evolved, the difference between its spectra and other less energetic BL-Ic events 
reduces. The spectrum of ASASSN-16fp obtained at $\sim$+5 day is overall very similar to the spectrum of 
SN~2002ap, SN~2009bb and SN~2012ap. All the spectral lines in ASASSN-16fp are broader than those in SN~2007gr
and narrower than SN~1998bw and SN~2014ad. This trend continues until the last spectrum presented here.
The features in the spectrum of ASASSN-16fp are better resolved than in other BL-Ic SNe and are marginally 
broader and blended than in SN~2004aw. The spectra of ASASSN-16fp at $\sim$+15 d and $\sim$+30 d continue to show
the strong feature at $\sim$6300 \AA\, identified with  Si\,{\sc ii} in the pre-maximum spectrum. In the 
spectra of other objects, the width and strength of this feature is relatively weak. The Na\,{\sc i} line in the BL-Ic 
SNe $\sim$+30 d shows broad absorption, while SN~2007gr and SN~2004aw show a sharp absorption feature. 
The width of Na\,{\sc i} line in ASASSN-16fp seems to lie in between the BL-Ic and transitional Type Ic 
SN~2004aw.    

\section{Summary and conclusions}\label{dis}

We have presented the results based on photometry (optical and {\it Swift} UV) and optical spectroscopy of the 
transitional Type Ic supernova ASASSN-16fp. The pre-maximum rise and early decline of the light curves are slow, 
making them broader in all the bands. There is no established co-relation between rise and decay time for SE-SNe. 
However, the light curve evolution of ASASSN-16fp follows a general trend of slow risers being also slow decliners 
\citep[see, also][]{2016MNRAS.458.2973P}.

The peak in $I$-band is significantly delayed ($\sim$7 d) with respect to $B$-band maximum, which is found to be
longer than the offset between $B$ and $I$ band maximum for Type Ic and BL-Ic SNe \citep[c.f.][]{2011ApJ...741...97D,
2015A&A...574A..60T}. In case of SN~2004aw the delay was $\sim$8 d \citep{2006MNRAS.371.1459T}.
It is noticed that light curve evolution of ASASSN-16fp and transitional Ic SN~2004aw \citep{2006MNRAS.371.1459T} 
exhibit similarities in all bands (cf. Figure~\ref{fig_lccomp}) including the similar $\Delta$m$_{15}$ values in
$BVRI$ bands (see Section~\ref{lc}).
In addition to ASASSN-16fp and SN~2004aw, there are just a handful of SE-SNe that have remarkably broad light 
curves with much longer rise times. The BL-Ic SN~1997ef \citep{2000ApJ...534..660I,2004ApJ...614..858M}, normal 
Type Ic SN~2011bm \citep{2012ApJ...749L..28V} and iPTF15dtg \citep{2016A&A...592A..89T} have a rise time of $\sim$35 d.
The long rise time in these events suggests large ejecta mass, indicating massive Wolf-Rayet stars as 
their progenitor. 

A combination of gradual cooling of the expanding ejecta and variation in the ionization states of different 
species result in flux change, which consequently appear in the form of temporal evolution of the colour.
The colour evolution of SE-SNe shows large dispersion (cf. Figure~\ref{fig_3}). The slow evolution and relatively 
redder $B-V$ colour of ASASSN-16fp suggests low photospheric temperature and/or slower cooling rate of the ejecta.
The absolute luminosity of SE-SNe also shows large variation, the BL-Ic are generally found to be brighter than the
normal Ic events \citep[c.f.][]{2016MNRAS.457..328L,2016MNRAS.458.2973P}. The absolute $V$-band luminosity of 
ASASSN-16fp lies towards the fainter end of the absolute luminosity distribution of BL-Ic SNe. Its luminosity is
comparable to transitional Type Ic SN~2004aw and brighter than normal Type Ic SN~2007gr.
A detailed discussion on the bulk photometric properties of SE SNe can be found elsewhere \citep[e.g.][]{2011ApJ...741...97D,
2014ApJS..213...19B,2015A&A...574A..60T,2016MNRAS.458.2973P}.
It appears that the light curve evolution pattern does not have direct impact on colour curves and absolute 
luminosity values. Nonetheless, such distribution points towards different progenitor channels and explosion 
mechanisms.

The spectral evolution of ASASSN-16fp exhibits a similarity with the transitional Type Ic SN~2004aw. The
width of the spectral lines of ASASSN-16fp are intermediate between transitional Type Ic SN~2004aw and BL-Ic 
SN~2002ap \citep{2002MNRAS.332L..73G,2002ApJ...572L..61M,2003PASP..115.1220F}. Most of the well developed 
spectral features were present in the first spectrum obtained $\sim$--9 d and were found to evolve slowly.
The initial photospheric velocity of ASASSN-16fp was moderately high ($\sim$20000 km s$^{-1}$), comparable
to SN~2007ru at similar epochs. In the pre-maximum phase, C\,{\sc i} lines were detected but with weaker 
strength than SN~2007gr. 

Recently, \citet{2017ApJ...837....1Y} have presented their results on ASASSN-16fp. They have identified 
He\,{\sc i} lines in their pre-maximum spectra and classified this event as broad-lined Type Ib SN. However, 
they also mentioned that in many respects ASASSN-16fp is similar to a BL-Ic SN (see their Section 3.1). 
The classification of SNe Ib and Ic (BL-Ic) is based on presence or absence of He\,{\sc i} lines in their 
spectra, though it is debatable. Some theoretical models predict that progenitors of Ic SNe may contain 
helium \citep[e.g.,][]{2012MNRAS.424.2139D}, but others argue in favor of helium-free SN Ic progenitors 
\citep[e.g.][]{2012MNRAS.422...70H,2013ApJ...773L...7F}. In some of the Ic and BL-Ic SNe, He\,{\sc i} 
lines have been detected e.g. SN~1994I \citep{1996ApJ...462..462C}, SN~2004aw \citep{2006MNRAS.371.1459T},
SN~2009bb \citep{2011ApJ...728...14P} and SN~2012ap \citep{2015ApJ...799...51M}. In case of SN Ib, the 
strength of the He\,{\sc i} lines increases with time \citep{1998MNRAS.295..428M,2001AJ....121.1648M}. 
\citet{2017ApJ...837....1Y} have shown that around 15 d after explosion, the He\,{\sc i} lines became very 
weak in ASASSN-16fp. The identification of He\,{\sc i} 20587 \AA\, in the NIR spectrum is another strong 
clue to confirm the presence of the helium \citep{2006MNRAS.371.1459T}. A better follow-up of NIR spectrum 
along with non-LTE spectral codes (to properly treat non-thermal excitations) could help identify the helium 
features more robustly, shedding some light on evolutionary stages of massive stars 
\citep[e.g.][]{2005A&A...443..643Y,2012ARA&A..50..107L}.

The light curve and spectral evolution of ASASSN-16fp indicate that it is a transitional Type Ic supernova 
(similar to SN~2004aw), having spectral properties between normal Type Ic and broad-lined Type Ic. The 
synthesized radioactive $^{56}$Ni powers the SN light curve. The $^{56}$Ni mass in BL-Ic/GRB-SNe is 
generally high \citep{2011ApJ...741...97D,2013MNRAS.434.1098C,2016MNRAS.457..328L,2016MNRAS.458.2973P}, 
making them more luminous than other SE-SNe.
However, it may also be possible that even if a similar amount of $^{56}$Ni is produced, the slow expanding 
SNe may be fainter \citep{2016LPICo1962.4116C}. Furthermore, the location and degree of mixing of nickel 
will also affect the observational features. The occurrence of late peak and slow evolution of the light
curve of ASASSN-16fp may possibly be due to the shallow distribution and deeply located $^{56}$Ni in the 
ejecta \citep[see][]{2013ApJ...769...67P}.

The explosion parameters of ASASSN-16fp were estimated by fitting Arnett-Valenti analytical model to the 
bolometric light curve (see Section~\ref{bol}). The derived value of mass of the ejecta ($M_{\rm ej}$) 
$\sim$4.5\,$\pm$\,0.3 M$_{\odot}$ and kinetic energy ($E_{\rm k}$) $\sim$6.9$^{+1.5}_{-1.3}$ $\times$ 
10$^{51}$ erg, are comparable to those of other BL-Ic events \citep{2016MNRAS.457..328L}.
However, mass of $^{56}$Ni ($M_{\rm Ni}$) 0.10 $\pm$ 0.01 M$_{\odot}$ for ASASSN-16fp is found to be 
lower than the median value estimated for BL-Ic SNe \citep{2013MNRAS.434.1098C,2016MNRAS.457..328L}.
We also estimated physical parameters of ASASSN-16fp using Vink{\'o}'s analytical model and the 
derived values were found to be consistent with those obtained from Arnett-Valenti model.

The observed diversity in Ic/BL-Ic SNe has been attributed to the asphericity of the ejecta and/or jet induced 
explosions \citep[e.g.][]{1999ApJ...524L.107K,2002ApJ...580L..39K}. The spectropolarimetric/polarimetric data 
will be helpful in constraining the geometry of the explosion.
The spectral evolution during the nebular phase will reveal the deeper layers of the ejecta. The line profile 
in the nebular phase also provides clues about the geometry of the ejecta.  
\citet{2005Sci...308.1284M} suggested that double-peaked emission-line profiles of [O\,{\sc i}] may be indicative 
of asymmetric geometry of the explosion. Such features have been found in many SNe \citep{2008Sci...319.1220M,
2008ApJ...687L...9M,2009MNRAS.397..677T,2010ApJ...709.1343M}. A critical analysis of the nebular phase spectra of
ASASSN-16fp will be useful for understanding the observed properties in greater detail.


\section*{Acknowledgments}
The authors thank the referee for his/her critical review and constructive suggestions that helped to 
improve the contents and presentation of the paper. 
BK acknowledges the Science and Engineering Research Board (SERB) under the Department of Science 
\& Technology, Govt. of India, for financial assistance in the form of National Post-Doctoral Fellowship 
(Ref. No. PDF/2016/001563).
We are grateful to the observers at HCT who provided their valuable time and support for the observations 
of this event. We thank T. P. Prabhu for various discussions during the preparation of this manuscript.
We also acknowledge the Weizmann interactive supernova data repository - \url{http://wiserep.weizmann.ac.il}.
This work has made use of the public data in the {\it Swift} data archive and the NASA/IPAC Extragalactic
Database (NED) which is operated by Jet Propulsion Laboratory, California Institute of Technology, under
contract with the National Aeronautics and Space Administration (NASA).

\bibliographystyle{mnras} 
\bibliography{sn16fp}

\begin{thebibliography}{}
\makeatletter
\relax
\def\mn@urlcharsother{\let\do\@makeother \do\$\do\&\do\#\do\^\do\_\do\%\do\~}
\def\mn@doi{\begingroup\mn@urlcharsother \@ifnextchar [ {\mn@doi@}
  {\mn@doi@[]}}
\def\mn@doi@[#1]#2{\def\@tempa{#1}\ifx\@tempa\@empty \href
  {http://dx.doi.org/#2} {doi:#2}\else \href {http://dx.doi.org/#2} {#1}\fi
  \endgroup}
\def\mn@eprint#1#2{\mn@eprint@#1:#2::\@nil}
\def\mn@eprint@arXiv#1{\href {http://arxiv.org/abs/#1} {{\tt arXiv:#1}}}
\def\mn@eprint@dblp#1{\href {http://dblp.uni-trier.de/rec/bibtex/#1.xml}
  {dblp:#1}}
\def\mn@eprint@#1:#2:#3:#4\@nil{\def\@tempa {#1}\def\@tempb {#2}\def\@tempc
  {#3}\ifx \@tempc \@empty \let \@tempc \@tempb \let \@tempb \@tempa \fi \ifx
  \@tempb \@empty \def\@tempb {arXiv}\fi \@ifundefined
  {mn@eprint@\@tempb}{\@tempb:\@tempc}{\expandafter \expandafter \csname
  mn@eprint@\@tempb\endcsname \expandafter{\@tempc}}}

\bibitem[\protect\citeauthoryear{{Argo}, {Romero-Canizales}, {Beswick}  \&
  {Prieto}}{{Argo} et~al.}{2016}]{2016ATel.9147....1A}
{Argo} M.~K.,  {Romero-Canizales} C.,  {Beswick} R.,   {Prieto} J.~L.,  2016,
  The Astronomer's Telegram, \href
  {http://adsabs.harvard.edu/abs/2016ATel.9147....1A} {9147}

\bibitem[\protect\citeauthoryear{{Arnett}}{{Arnett}}{1982}]{1982ApJ...253..785A}
{Arnett} W.~D.,  1982, \mn@doi [\apj] {10.1086/159681}, \href
  {http://adsabs.harvard.edu/abs/1982ApJ...253..785A} {253, 785}

\bibitem[\protect\citeauthoryear{{Arnett}}{{Arnett}}{1996}]{1996snih.book.....A}
{Arnett} D.,  1996, Supernovae and Nucleosynthesis: An Investigation of the
  History of Matter from the Big Bang to the Present. Princeton Univ. Press,
  Princeton, NJ

\bibitem[\protect\citeauthoryear{{Barbon}, {Benetti}, {Rosino}, {Cappellaro}
  \& {Turatto}}{{Barbon} et~al.}{1990}]{1990A&A...237...79B}
{Barbon} R.,  {Benetti} S.,  {Rosino} L.,  {Cappellaro} E.,   {Turatto} M.,
  1990, \aap, \href {http://adsabs.harvard.edu/abs/1990A%26A...237...79B} {237,
  79}

\bibitem[\protect\citeauthoryear{{Bessell}, {Castelli}  \& {Plez}}{{Bessell}
  et~al.}{1998}]{1998A&A...333..231B}
{Bessell} M.~S.,  {Castelli} F.,   {Plez} B.,  1998, \aap, \href
  {http://adsabs.harvard.edu/abs/1998A%26A...333..231B} {333, 231}

\bibitem[\protect\citeauthoryear{{Bianco} et~al.,}{{Bianco}
  et~al.}{2014}]{2014ApJS..213...19B}
{Bianco} F.~B.,  et~al., 2014, \mn@doi [\apjs] {10.1088/0067-0049/213/2/19},
  \href {http://adsabs.harvard.edu/abs/2014ApJS..213...19B} {213, 19}

\bibitem[\protect\citeauthoryear{{Branch}, {Jeffery}, {Young}  \&
  {Baron}}{{Branch} et~al.}{2006}]{2006PASP..118..791B}
{Branch} D.,  {Jeffery} D.~J.,  {Young} T.~R.,   {Baron} E.,  2006, \mn@doi
  [\pasp] {10.1086/505548}, \href
  {http://adsabs.harvard.edu/abs/2006PASP..118..791B} {118, 791}

\bibitem[\protect\citeauthoryear{{Brown} et~al.,}{{Brown}
  et~al.}{2009}]{2009AJ....137.4517B}
{Brown} P.~J.,  et~al., 2009, \mn@doi [\aj] {10.1088/0004-6256/137/5/4517},
  \href {http://cdsads.u-strasbg.fr/abs/2009AJ....137.4517B} {137, 4517}

\bibitem[\protect\citeauthoryear{{Cano}}{{Cano}}{2013}]{2013MNRAS.434.1098C}
{Cano} Z.,  2013, \mn@doi [\mnras] {10.1093/mnras/stt1048}, \href
  {http://cdsads.u-strasbg.fr/abs/2013MNRAS.434.1098C} {434, 1098}

\bibitem[\protect\citeauthoryear{{Cano}}{{Cano}}{2016}]{2016LPICo1962.4116C}
{Cano} Z.,  2016, LPI Contributions, \href
  {http://adsabs.harvard.edu/abs/2016LPICo1962.4116C} {1962, 4116}

\bibitem[\protect\citeauthoryear{{Cano} et~al.,}{{Cano}
  et~al.}{2011}]{2011ApJ...740...41C}
{Cano} Z.,  et~al., 2011, \mn@doi [\apj] {10.1088/0004-637X/740/1/41}, \href
  {http://adsabs.harvard.edu/abs/2011ApJ...740...41C} {740, 41}

\bibitem[\protect\citeauthoryear{{Chugai}}{{Chugai}}{2000}]{2000AstL...26..797C}
{Chugai} N.~N.,  2000, \mn@doi [Astronomy Letters] {10.1134/1.1331160}, \href
  {http://adsabs.harvard.edu/abs/2000AstL...26..797C} {26, 797}

\bibitem[\protect\citeauthoryear{{Clocchiatti} \& {Wheeler}}{{Clocchiatti} \&
  {Wheeler}}{1997}]{1997ApJ...491..375C}
{Clocchiatti} A.,  {Wheeler} J.~C.,  1997, \apj, \href
  {http://adsabs.harvard.edu/abs/1997ApJ...491..375C} {491, 375}

\bibitem[\protect\citeauthoryear{{Clocchiatti}, {Wheeler}, {Brotherton},
  {Cochran}, {Wills}, {Barker}  \& {Turatto}}{{Clocchiatti}
  et~al.}{1996}]{1996ApJ...462..462C}
{Clocchiatti} A.,  {Wheeler} J.~C.,  {Brotherton} M.~S.,  {Cochran} A.~L.,
  {Wills} D.,  {Barker} E.~S.,   {Turatto} M.,  1996, \mn@doi [\apj]
  {10.1086/177165}, \href {http://adsabs.harvard.edu/abs/1996ApJ...462..462C}
  {462, 462}

\bibitem[\protect\citeauthoryear{{Clocchiatti}, {Suntzeff}, {Covarrubias}  \&
  {Candia}}{{Clocchiatti} et~al.}{2011}]{2011AJ....141..163C}
{Clocchiatti} A.,  {Suntzeff} N.~B.,  {Covarrubias} R.,   {Candia} P.,  2011,
  \mn@doi [\aj] {10.1088/0004-6256/141/5/163}, \href
  {http://adsabs.harvard.edu/abs/2011AJ....141..163C} {141, 163}

\bibitem[\protect\citeauthoryear{{Dessart}, {Hillier}, {Li}  \&
  {Woosley}}{{Dessart} et~al.}{2012}]{2012MNRAS.424.2139D}
{Dessart} L.,  {Hillier} D.~J.,  {Li} C.,   {Woosley} S.,  2012, \mn@doi
  [\mnras] {10.1111/j.1365-2966.2012.21374.x}, \href
  {http://adsabs.harvard.edu/abs/2012MNRAS.424.2139D} {424, 2139}

\bibitem[\protect\citeauthoryear{{Dessart}, {Hillier}, {Woosley}, {Livne},
  {Waldman}, {Yoon}  \& {Langer}}{{Dessart} et~al.}{2016}]{2016MNRAS.458.1618D}
{Dessart} L.,  {Hillier} D.~J.,  {Woosley} S.,  {Livne} E.,  {Waldman} R.,
  {Yoon} S.-C.,   {Langer} N.,  2016, \mn@doi [\mnras] {10.1093/mnras/stw418},
  \href {http://adsabs.harvard.edu/abs/2016MNRAS.458.1618D} {458, 1618}

\bibitem[\protect\citeauthoryear{{Drout} et~al.,}{{Drout}
  et~al.}{2011}]{2011ApJ...741...97D}
{Drout} M.~R.,  et~al., 2011, \mn@doi [\apj] {10.1088/0004-637X/741/2/97},
  \href {http://adsabs.harvard.edu/abs/2011ApJ...741...97D} {741, 97}

\bibitem[\protect\citeauthoryear{{Drout} et~al.,}{{Drout}
  et~al.}{2016}]{2016ApJ...821...57D}
{Drout} M.~R.,  et~al., 2016, \mn@doi [\apj] {10.3847/0004-637X/821/1/57},
  \href {http://adsabs.harvard.edu/abs/2016ApJ...821...57D} {821, 57}

\bibitem[\protect\citeauthoryear{{Elias-Rosa} et~al.,}{{Elias-Rosa}
  et~al.}{2016}]{2016ATel.9090....1E}
{Elias-Rosa} N.,  et~al., 2016, The Astronomer's Telegram, \href
  {http://adsabs.harvard.edu/abs/2016ATel.9090....1E} {9090}

\bibitem[\protect\citeauthoryear{{Elmhamdi}, {Danziger}, {Branch},
  {Leibundgut}, {Baron}  \& {Kirshner}}{{Elmhamdi}
  et~al.}{2006}]{2006A&A...450..305E}
{Elmhamdi} A.,  {Danziger} I.~J.,  {Branch} D.,  {Leibundgut} B.,  {Baron} E.,
   {Kirshner} R.~P.,  2006, \mn@doi [\aap] {10.1051/0004-6361:20054366}, \href
  {http://adsabs.harvard.edu/abs/2006A%26A...450..305E} {450, 305}

\bibitem[\protect\citeauthoryear{{Filippenko}}{{Filippenko}}{1997}]{1997ARA&A..35..309F}
{Filippenko} A.~V.,  1997, \mn@doi [\araa] {10.1146/annurev.astro.35.1.309},
  \href {http://adsabs.harvard.edu/abs/1997ARA%26A..35..309F} {35, 309}

\bibitem[\protect\citeauthoryear{{Foley} et~al.,}{{Foley}
  et~al.}{2003}]{2003PASP..115.1220F}
{Foley} R.~J.,  et~al., 2003, \mn@doi [\pasp] {10.1086/378242}, \href
  {http://adsabs.harvard.edu/abs/2003PASP..115.1220F} {115, 1220}

\bibitem[\protect\citeauthoryear{{Frey}, {Fryer}  \& {Young}}{{Frey}
  et~al.}{2013}]{2013ApJ...773L...7F}
{Frey} L.~H.,  {Fryer} C.~L.,   {Young} P.~A.,  2013, \mn@doi [\apjl]
  {10.1088/2041-8205/773/1/L7}, \href
  {http://adsabs.harvard.edu/abs/2013ApJ...773L...7F} {773, L7}

\bibitem[\protect\citeauthoryear{{Gal-Yam}, {Ofek}  \& {Shemmer}}{{Gal-Yam}
  et~al.}{2002}]{2002MNRAS.332L..73G}
{Gal-Yam} A.,  {Ofek} E.~O.,   {Shemmer} O.,  2002, \mn@doi [\mnras]
  {10.1046/j.1365-8711.2002.05535.x}, \href
  {http://adsabs.harvard.edu/abs/2002MNRAS.332L..73G} {332, L73}

\bibitem[\protect\citeauthoryear{{Galama} et~al.,}{{Galama}
  et~al.}{1998}]{1998Natur.395..670G}
{Galama} T.~J.,  et~al., 1998, \mn@doi [\nat] {10.1038/27150}, \href
  {http://adsabs.harvard.edu/abs/1998Natur.395..670G} {395, 670}

\bibitem[\protect\citeauthoryear{{Grupe}, {Brown}, {Dong}, {Shappee},
  {Holoien}, {Stanek}, {Prieto}  \& {Margutti}}{{Grupe}
  et~al.}{2016}]{2016ATel.9088....1G}
{Grupe} D.,  {Brown} P.,  {Dong} S.,  {Shappee} B.~J.,  {Holoien} T.,  {Stanek}
  K.,  {Prieto} J.~L.,   {Margutti} R.,  2016, The Astronomer's Telegram, \href
  {http://adsabs.harvard.edu/abs/2016ATel.9088....1G} {9088}

\bibitem[\protect\citeauthoryear{{Hachinger}, {Mazzali}, {Taubenberger},
  {Hillebrandt}, {Nomoto}  \& {Sauer}}{{Hachinger}
  et~al.}{2012}]{2012MNRAS.422...70H}
{Hachinger} S.,  {Mazzali} P.~A.,  {Taubenberger} S.,  {Hillebrandt} W.,
  {Nomoto} K.,   {Sauer} D.~N.,  2012, \mn@doi [\mnras]
  {10.1111/j.1365-2966.2012.20464.x}, \href
  {http://adsabs.harvard.edu/abs/2012MNRAS.422...70H} {422, 70}

\bibitem[\protect\citeauthoryear{{Heger}, {Fryer}, {Woosley}, {Langer}  \&
  {Hartmann}}{{Heger} et~al.}{2003}]{2003ApJ...591..288H}
{Heger} A.,  {Fryer} C.~L.,  {Woosley} S.~E.,  {Langer} N.,   {Hartmann} D.~H.,
   2003, \mn@doi [\apj] {10.1086/375341}, \href
  {http://adsabs.harvard.edu/abs/2003ApJ...591..288H} {591, 288}

\bibitem[\protect\citeauthoryear{{Holoien} et~al.,}{{Holoien}
  et~al.}{2016}]{2016ATel.9086....1H}
{Holoien} T.~W.-S.,  et~al., 2016, The Astronomer's Telegram, \href
  {http://adsabs.harvard.edu/abs/2016ATel.9086....1H} {9086}

\bibitem[\protect\citeauthoryear{{Hunter} et~al.,}{{Hunter}
  et~al.}{2009}]{2009A&A...508..371H}
{Hunter} D.~J.,  et~al., 2009, \mn@doi [\aap] {10.1051/0004-6361/200912896},
  \href {http://adsabs.harvard.edu/abs/2009A%26A...508..371H} {508, 371}

\bibitem[\protect\citeauthoryear{{Iwamoto} et~al.,}{{Iwamoto}
  et~al.}{1998}]{1998Natur.395..672I}
{Iwamoto} K.,  et~al., 1998, \mn@doi [\nat] {10.1038/27155}, \href
  {http://adsabs.harvard.edu/abs/1998Natur.395..672I} {395, 672}

\bibitem[\protect\citeauthoryear{{Iwamoto} et~al.,}{{Iwamoto}
  et~al.}{2000}]{2000ApJ...534..660I}
{Iwamoto} K.,  et~al., 2000, \mn@doi [\apj] {10.1086/308761}, \href
  {http://adsabs.harvard.edu/abs/2000ApJ...534..660I} {534, 660}

\bibitem[\protect\citeauthoryear{{Kawabata} et~al.,}{{Kawabata}
  et~al.}{2002}]{2002ApJ...580L..39K}
{Kawabata} K.~S.,  et~al., 2002, \mn@doi [\apjl] {10.1086/345545}, \href
  {http://adsabs.harvard.edu/abs/2002ApJ...580L..39K} {580, L39}

\bibitem[\protect\citeauthoryear{{Khokhlov}, {H{\"o}flich}, {Oran}, {Wheeler},
  {Wang}  \& {Chtchelkanova}}{{Khokhlov} et~al.}{1999}]{1999ApJ...524L.107K}
{Khokhlov} A.~M.,  {H{\"o}flich} P.~A.,  {Oran} E.~S.,  {Wheeler} J.~C.,
  {Wang} L.,   {Chtchelkanova} A.~Y.,  1999, \mn@doi [\apjl] {10.1086/312305},
  \href {http://adsabs.harvard.edu/abs/1999ApJ...524L.107K} {524, L107}

\bibitem[\protect\citeauthoryear{{Landolt}}{{Landolt}}{1992}]{1992AJ....104..340L}
{Landolt} A.~U.,  1992, \mn@doi [\aj] {10.1086/116242}, \href
  {http://adsabs.harvard.edu/abs/1992AJ....104..340L} {104, 340}

\bibitem[\protect\citeauthoryear{{Langer}}{{Langer}}{2012}]{2012ARA&A..50..107L}
{Langer} N.,  2012, \mn@doi [\araa] {10.1146/annurev-astro-081811-125534},
  \href {http://adsabs.harvard.edu/abs/2012ARA%26A..50..107L} {50, 107}

\bibitem[\protect\citeauthoryear{{Liu}, {Modjaz}, {Bianco}  \& {Graur}}{{Liu}
  et~al.}{2016}]{2016APJ-637X-827-2-90}
{Liu} Y.-Q.,  {Modjaz} M.,  {Bianco} F.~B.,   {Graur} O.,  2016, \apj, \href
  {http://stacks.iop.org/0004-637X/827/i=2/a=90} {827, 90}

\bibitem[\protect\citeauthoryear{{Lyman}, {Bersier}, {James}, {Mazzali},
  {Eldridge}, {Fraser}  \& {Pian}}{{Lyman} et~al.}{2016}]{2016MNRAS.457..328L}
{Lyman} J.~D.,  {Bersier} D.,  {James} P.~A.,  {Mazzali} P.~A.,  {Eldridge}
  J.~J.,  {Fraser} M.,   {Pian} E.,  2016, \mn@doi [\mnras]
  {10.1093/mnras/stv2983}, \href
  {http://adsabs.harvard.edu/abs/2016MNRAS.457..328L} {457, 328}

\bibitem[\protect\citeauthoryear{{Maeda} et~al.,}{{Maeda}
  et~al.}{2008}]{2008Sci...319.1220M}
{Maeda} K.,  et~al., 2008, \mn@doi [Science] {10.1126/science.1149437}, \href
  {http://adsabs.harvard.edu/abs/2008Sci...319.1220M} {319, 1220}

\bibitem[\protect\citeauthoryear{{Matheson}, {Filippenko}, {Li}, {Leonard}  \&
  {Shields}}{{Matheson} et~al.}{2001}]{2001AJ....121.1648M}
{Matheson} T.,  {Filippenko} A.~V.,  {Li} W.,  {Leonard} D.~C.,   {Shields}
  J.~C.,  2001, \mn@doi [\aj] {10.1086/319390}, \href
  {http://adsabs.harvard.edu/abs/2001AJ....121.1648M} {121, 1648}

\bibitem[\protect\citeauthoryear{{Mazzali} \& {Lucy}}{{Mazzali} \&
  {Lucy}}{1998}]{1998MNRAS.295..428M}
{Mazzali} P.~A.,  {Lucy} L.~B.,  1998, \mn@doi [\mnras]
  {10.1046/j.1365-8711.1998.01323.x}, \href
  {http://adsabs.harvard.edu/abs/1998MNRAS.295..428M} {295, 428}

\bibitem[\protect\citeauthoryear{{Mazzali}, {Iwamoto}  \& {Nomoto}}{{Mazzali}
  et~al.}{2000}]{2000ApJ...545..407M}
{Mazzali} P.~A.,  {Iwamoto} K.,   {Nomoto} K.,  2000, \mn@doi [\apj]
  {10.1086/317808}, \href {http://adsabs.harvard.edu/abs/2000ApJ...545..407M}
  {545, 407}

\bibitem[\protect\citeauthoryear{{Mazzali} et~al.,}{{Mazzali}
  et~al.}{2002}]{2002ApJ...572L..61M}
{Mazzali} P.~A.,  et~al., 2002, \mn@doi [\apjl] {10.1086/341504}, \href
  {http://adsabs.harvard.edu/abs/2002ApJ...572L..61M} {572, L61}

\bibitem[\protect\citeauthoryear{{Mazzali}, {Deng}, {Maeda}, {Nomoto},
  {Filippenko}  \& {Matheson}}{{Mazzali} et~al.}{2004}]{2004ApJ...614..858M}
{Mazzali} P.~A.,  {Deng} J.,  {Maeda} K.,  {Nomoto} K.,  {Filippenko} A.~V.,
  {Matheson} T.,  2004, \mn@doi [\apj] {10.1086/423888}, \href
  {http://adsabs.harvard.edu/abs/2004ApJ...614..858M} {614, 858}

\bibitem[\protect\citeauthoryear{{Mazzali} et~al.,}{{Mazzali}
  et~al.}{2005}]{2005Sci...308.1284M}
{Mazzali} P.~A.,  et~al., 2005, \mn@doi [Science] {10.1126/science.1111384},
  \href {http://adsabs.harvard.edu/abs/2005Sci...308.1284M} {308, 1284}

\bibitem[\protect\citeauthoryear{{Milisavljevic}, {Fesen}, {Gerardy},
  {Kirshner}  \& {Challis}}{{Milisavljevic} et~al.}{2010}]{2010ApJ...709.1343M}
{Milisavljevic} D.,  {Fesen} R.~A.,  {Gerardy} C.~L.,  {Kirshner} R.~P.,
  {Challis} P.,  2010, \mn@doi [\apj] {10.1088/0004-637X/709/2/1343}, \href
  {http://adsabs.harvard.edu/abs/2010ApJ...709.1343M} {709, 1343}

\bibitem[\protect\citeauthoryear{{Milisavljevic} et~al.,}{{Milisavljevic}
  et~al.}{2015}]{2015ApJ...799...51M}
{Milisavljevic} D.,  et~al., 2015, \mn@doi [\apj] {10.1088/0004-637X/799/1/51},
  \href {http://adsabs.harvard.edu/abs/2015ApJ...799...51M} {799, 51}

\bibitem[\protect\citeauthoryear{{Minkowski}}{{Minkowski}}{1941}]{1941PASP...53..224M}
{Minkowski} R.,  1941, \mn@doi [\pasp] {10.1086/125315}, \href
  {http://adsabs.harvard.edu/abs/1941PASP...53..224M} {53, 224}

\bibitem[\protect\citeauthoryear{{Mirabal}, {Halpern}, {An}, {Thorstensen}  \&
  {Terndrup}}{{Mirabal} et~al.}{2006}]{2006ApJ...643L..99M}
{Mirabal} N.,  {Halpern} J.~P.,  {An} D.,  {Thorstensen} J.~R.,   {Terndrup}
  D.~M.,  2006, \mn@doi [\apjl] {10.1086/505177}, \href
  {http://adsabs.harvard.edu/abs/2006ApJ...643L..99M} {643, L99}

\bibitem[\protect\citeauthoryear{{Modjaz}, {Kirshner}, {Blondin}, {Challis}  \&
  {Matheson}}{{Modjaz} et~al.}{2008}]{2008ApJ...687L...9M}
{Modjaz} M.,  {Kirshner} R.~P.,  {Blondin} S.,  {Challis} P.,   {Matheson} T.,
  2008, \mn@doi [\apjl] {10.1086/593135}, \href
  {http://adsabs.harvard.edu/abs/2008ApJ...687L...9M} {687, L9}

\bibitem[\protect\citeauthoryear{{Modjaz} et~al.,}{{Modjaz}
  et~al.}{2014}]{2014AJ....147...99M}
{Modjaz} M.,  et~al., 2014, \mn@doi [\aj] {10.1088/0004-6256/147/5/99}, \href
  {http://adsabs.harvard.edu/abs/2014AJ....147...99M} {147, 99}

\bibitem[\protect\citeauthoryear{{Mooley} et~al.,}{{Mooley}
  et~al.}{2016}]{2016ATel.9134....1M}
{Mooley} K.~P.,  et~al., 2016, The Astronomer's Telegram, \href
  {http://adsabs.harvard.edu/abs/2016ATel.9134....1M} {9134}

\bibitem[\protect\citeauthoryear{{Nayana} \& {Chandra}}{{Nayana} \&
  {Chandra}}{2016}]{2016ATel.9201....1N}
{Nayana} A.~J.,  {Chandra} P.,  2016, The Astronomer's Telegram, \href
  {http://adsabs.harvard.edu/abs/2016ATel.9201....1N} {9201}

\bibitem[\protect\citeauthoryear{{Olivares E.} et~al.,}{{Olivares E.}
  et~al.}{2012}]{2012A&A...539A..76O}
{Olivares E.} F.,  et~al., 2012, \mn@doi [\aap] {10.1051/0004-6361/201117929},
  \href {http://adsabs.harvard.edu/abs/2012A%26A...539A..76O} {539, A76}

\bibitem[\protect\citeauthoryear{{Pandey}, {Anupama}, {Sagar}, {Bhattacharya},
  {Sahu}  \& {Pandey}}{{Pandey} et~al.}{2003}]{2003MNRAS.340..375P}
{Pandey} S.~B.,  {Anupama} G.~C.,  {Sagar} R.,  {Bhattacharya} D.,  {Sahu}
  D.~K.,   {Pandey} J.~C.,  2003, \mn@doi [\mnras]
  {10.1046/j.1365-8711.2003.06148.x}, \href
  {http://adsabs.harvard.edu/abs/2003MNRAS.340..375P} {340, 375}

\bibitem[\protect\citeauthoryear{{Parrent}, {Milisavljevic}, {Soderberg}  \&
  {Parthasarathy}}{{Parrent} et~al.}{2016}]{2016ApJ...820...75P}
{Parrent} J.~T.,  {Milisavljevic} D.,  {Soderberg} A.~M.,   {Parthasarathy} M.,
   2016, \mn@doi [\apj] {10.3847/0004-637X/820/1/75}, \href
  {http://adsabs.harvard.edu/abs/2016ApJ...820...75P} {820, 75}

\bibitem[\protect\citeauthoryear{{Patat} et~al.,}{{Patat}
  et~al.}{2001}]{2001ApJ...555..900P}
{Patat} F.,  et~al., 2001, \mn@doi [\apj] {10.1086/321526}, \href
  {http://adsabs.harvard.edu/abs/2001ApJ...555..900P} {555, 900}

\bibitem[\protect\citeauthoryear{{Pian} et~al.,}{{Pian}
  et~al.}{2006}]{2006Natur.442.1011P}
{Pian} E.,  et~al., 2006, \mn@doi [\nat] {10.1038/nature05082}, \href
  {http://cdsads.u-strasbg.fr/abs/2006Natur.442.1011P} {442, 1011}

\bibitem[\protect\citeauthoryear{{Pignata} et~al.,}{{Pignata}
  et~al.}{2011}]{2011ApJ...728...14P}
{Pignata} G.,  et~al., 2011, \mn@doi [\apj] {10.1088/0004-637X/728/1/14}, \href
  {http://adsabs.harvard.edu/abs/2011ApJ...728...14P} {728, 14}

\bibitem[\protect\citeauthoryear{{Piro} \& {Nakar}}{{Piro} \&
  {Nakar}}{2013}]{2013ApJ...769...67P}
{Piro} A.~L.,  {Nakar} E.,  2013, \mn@doi [\apj] {10.1088/0004-637X/769/1/67},
  \href {http://adsabs.harvard.edu/abs/2013ApJ...769...67P} {769, 67}

\bibitem[\protect\citeauthoryear{{Podsiadlowski}, {Joss}  \&
  {Hsu}}{{Podsiadlowski} et~al.}{1992}]{1992ApJ...391..246P}
{Podsiadlowski} P.,  {Joss} P.~C.,   {Hsu} J.~J.~L.,  1992, \mn@doi [\apj]
  {10.1086/171341}, \href {http://adsabs.harvard.edu/abs/1992ApJ...391..246P}
  {391, 246}

\bibitem[\protect\citeauthoryear{{Poole} et~al.,}{{Poole}
  et~al.}{2008}]{2008MNRAS.383..627P}
{Poole} T.~S.,  et~al., 2008, \mn@doi [\mnras]
  {10.1111/j.1365-2966.2007.12563.x}, \href
  {http://adsabs.harvard.edu/abs/2008MNRAS.383..627P} {383, 627}

\bibitem[\protect\citeauthoryear{{Poznanski}, {Ganeshalingam}, {Silverman}  \&
  {Filippenko}}{{Poznanski} et~al.}{2011}]{2011MNRAS.415L..81P}
{Poznanski} D.,  {Ganeshalingam} M.,  {Silverman} J.~M.,   {Filippenko} A.~V.,
  2011, \mn@doi [\mnras] {10.1111/j.1745-3933.2011.01084.x}, \href
  {http://adsabs.harvard.edu/abs/2011MNRAS.415L..81P} {415, L81}

\bibitem[\protect\citeauthoryear{{Prentice} et~al.,}{{Prentice}
  et~al.}{2016}]{2016MNRAS.458.2973P}
{Prentice} S.~J.,  et~al., 2016, \mn@doi [\mnras] {10.1093/mnras/stw299}, \href
  {http://adsabs.harvard.edu/abs/2016MNRAS.458.2973P} {458, 2973}

\bibitem[\protect\citeauthoryear{{Pritchard}, {Roming}, {Brown}, {Bayless}  \&
  {Frey}}{{Pritchard} et~al.}{2014}]{2014ApJ...787..157P}
{Pritchard} T.~A.,  {Roming} P.~W.~A.,  {Brown} P.~J.,  {Bayless} A.~J.,
  {Frey} L.~H.,  2014, \mn@doi [\apj] {10.1088/0004-637X/787/2/157}, \href
  {http://cdsads.u-strasbg.fr/abs/2014ApJ...787..157P} {787, 157}

\bibitem[\protect\citeauthoryear{{Puls}, {Vink}  \& {Najarro}}{{Puls}
  et~al.}{2008}]{2008A&ARv..16..209P}
{Puls} J.,  {Vink} J.~S.,   {Najarro} F.,  2008, \mn@doi [\aapr]
  {10.1007/s00159-008-0015-8}, \href
  {http://adsabs.harvard.edu/abs/2008A%26ARv..16..209P} {16, 209}

\bibitem[\protect\citeauthoryear{{Richardson}, {Branch}  \&
  {Baron}}{{Richardson} et~al.}{2006}]{2006AJ....131.2233R}
{Richardson} D.,  {Branch} D.,   {Baron} E.,  2006, \mn@doi [\aj]
  {10.1086/500578}, \href {http://adsabs.harvard.edu/abs/2006AJ....131.2233R}
  {131, 2233}

\bibitem[\protect\citeauthoryear{{Richardson}, {Jenkins}, {Wright}  \&
  {Maddox}}{{Richardson} et~al.}{2014}]{2014AJ....147..118R}
{Richardson} D.,  {Jenkins} III R.~L.,  {Wright} J.,   {Maddox} L.,  2014,
  \mn@doi [\aj] {10.1088/0004-6256/147/5/118}, \href
  {http://adsabs.harvard.edu/abs/2014AJ....147..118R} {147, 118}

\bibitem[\protect\citeauthoryear{{Sahu}, {Tanaka}, {Anupama}, {Gurugubelli}  \&
  {Nomoto}}{{Sahu} et~al.}{2009}]{2009ApJ...697..676S}
{Sahu} D.~K.,  {Tanaka} M.,  {Anupama} G.~C.,  {Gurugubelli} U.~K.,   {Nomoto}
  K.,  2009, \mn@doi [\apj] {10.1088/0004-637X/697/1/676}, \href
  {http://adsabs.harvard.edu/abs/2009ApJ...697..676S} {697, 676}

\bibitem[\protect\citeauthoryear{{Sahu}, {et}  \& {al.}}{{Sahu}
  et~al.}{2017}]{sahu_sn14ad}
{Sahu} D.~K.,  {et}  {al.} 2017, MNRAS, submitted

\bibitem[\protect\citeauthoryear{{Schlafly} \& {Finkbeiner}}{{Schlafly} \&
  {Finkbeiner}}{2011}]{2011ApJ...737..103S}
{Schlafly} E.~F.,  {Finkbeiner} D.~P.,  2011, \mn@doi [\apj]
  {10.1088/0004-637X/737/2/103}, \href
  {http://adsabs.harvard.edu/abs/2011ApJ...737..103S} {737, 103}

\bibitem[\protect\citeauthoryear{{Smartt}}{{Smartt}}{2009}]{2009ARA&A..47...63S}
{Smartt} S.~J.,  2009, \mn@doi [\araa] {10.1146/annurev-astro-082708-101737},
  \href {http://adsabs.harvard.edu/abs/2009ARA%26A..47...63S} {47, 63}

\bibitem[\protect\citeauthoryear{{Smith} \& {Owocki}}{{Smith} \&
  {Owocki}}{2006}]{2006ApJ...645L..45S}
{Smith} N.,  {Owocki} S.~P.,  2006, \mn@doi [\apjl] {10.1086/506523}, \href
  {http://adsabs.harvard.edu/abs/2006ApJ...645L..45S} {645, L45}

\bibitem[\protect\citeauthoryear{{Sollerman} et~al.,}{{Sollerman}
  et~al.}{2006}]{2006A&A...454..503S}
{Sollerman} J.,  et~al., 2006, \mn@doi [\aap] {10.1051/0004-6361:20065226},
  \href {http://adsabs.harvard.edu/abs/2006A%26A...454..503S} {454, 503}

\bibitem[\protect\citeauthoryear{{Stalin}, {Hegde}, {Sahu}, {Parihar},
  {Anupama}, {Bhatt}  \& {Prabhu}}{{Stalin} et~al.}{2008}]{2008BASI...36..111S}
{Stalin} C.~S.,  {Hegde} M.,  {Sahu} D.~K.,  {Parihar} P.~S.,  {Anupama} G.~C.,
   {Bhatt} B.~C.,   {Prabhu} T.~P.,  2008, in Astron. Soc. India Conf. Ser.,
  \href {http://adsabs.harvard.edu/abs/2008BASI...36..111S} {36, 111}

\bibitem[\protect\citeauthoryear{{Stetson}}{{Stetson}}{1987}]{1987PASP...99..191S}
{Stetson} P.~B.,  1987, \mn@doi [\pasp] {10.1086/131977}, \href
  {http://adsabs.harvard.edu/abs/1987PASP...99..191S} {99, 191}

\bibitem[\protect\citeauthoryear{{Stetson}}{{Stetson}}{1992}]{1992ASPC...25..297S}
{Stetson} P.~B.,  1992, in {Worrall} D.~M.,  {Biemesderfer} C.,   {Barnes} J.,
  eds,  ASP Conf. Ser. Vol. 25, Astronomical Data Analysis Software and Systems
  I. Astron. Soc. Pac., San Francisco. p.~297

\bibitem[\protect\citeauthoryear{{Taddia} et~al.,}{{Taddia}
  et~al.}{2015}]{2015A&A...574A..60T}
{Taddia} F.,  et~al., 2015, \mn@doi [\aap] {10.1051/0004-6361/201423915}, \href
  {http://adsabs.harvard.edu/abs/2015A%26A...574A..60T} {574, A60}

\bibitem[\protect\citeauthoryear{{Taddia} et~al.,}{{Taddia}
  et~al.}{2016}]{2016A&A...592A..89T}
{Taddia} F.,  et~al., 2016, \mn@doi [\aap] {10.1051/0004-6361/201628703}, \href
  {http://adsabs.harvard.edu/abs/2016A%26A...592A..89T} {592, A89}

\bibitem[\protect\citeauthoryear{{Taubenberger} et~al.,}{{Taubenberger}
  et~al.}{2006}]{2006MNRAS.371.1459T}
{Taubenberger} S.,  et~al., 2006, \mn@doi [\mnras]
  {10.1111/j.1365-2966.2006.10776.x}, \href
  {http://adsabs.harvard.edu/abs/2006MNRAS.371.1459T} {371, 1459}

\bibitem[\protect\citeauthoryear{{Taubenberger} et~al.,}{{Taubenberger}
  et~al.}{2009}]{2009MNRAS.397..677T}
{Taubenberger} S.,  et~al., 2009, \mn@doi [\mnras]
  {10.1111/j.1365-2966.2009.15003.x}, \href
  {http://adsabs.harvard.edu/abs/2009MNRAS.397..677T} {397, 677}

\bibitem[\protect\citeauthoryear{{Tomita} et~al.,}{{Tomita}
  et~al.}{2006}]{2006ApJ...644..400T}
{Tomita} H.,  et~al., 2006, \mn@doi [\apj] {10.1086/503554}, \href
  {http://adsabs.harvard.edu/abs/2006ApJ...644..400T} {644, 400}

\bibitem[\protect\citeauthoryear{{Toy} et~al.,}{{Toy}
  et~al.}{2016}]{2016ApJ...818...79T}
{Toy} V.~L.,  et~al., 2016, \mn@doi [\apj] {10.3847/0004-637X/818/1/79}, \href
  {http://adsabs.harvard.edu/abs/2016ApJ...818...79T} {818, 79}

\bibitem[\protect\citeauthoryear{{Turatto}, {Benetti}  \&
  {Cappellaro}}{{Turatto} et~al.}{2003}]{2003fthp.conf..200T}
{Turatto} M.,  {Benetti} S.,   {Cappellaro} E.,  2003, in {Hillebrandt} W.,
  {Leibundgut} B.,  eds, Proc. ESO-MPA-MPE Workshop, From Twilight to
  Highlight: The Physics of Supernovae. Springer, Berlin, p. 200.

\bibitem[\protect\citeauthoryear{{Valenti} et~al.,}{{Valenti}
  et~al.}{2008a}]{2008MNRAS.383.1485V}
{Valenti} S.,  et~al., 2008a, \mn@doi [\mnras]
  {10.1111/j.1365-2966.2007.12647.x}, \href
  {http://adsabs.harvard.edu/abs/2008MNRAS.383.1485V} {383, 1485}

\bibitem[\protect\citeauthoryear{{Valenti} et~al.,}{{Valenti}
  et~al.}{2008b}]{2008ApJ...673L.155V}
{Valenti} S.,  et~al., 2008b, \mn@doi [\apjl] {10.1086/527672}, \href
  {http://adsabs.harvard.edu/abs/2008ApJ...673L.155V} {673, L155}

\bibitem[\protect\citeauthoryear{{Valenti} et~al.,}{{Valenti}
  et~al.}{2012}]{2012ApJ...749L..28V}
{Valenti} S.,  et~al., 2012, \mn@doi [\apjl] {10.1088/2041-8205/749/2/L28},
  \href {http://adsabs.harvard.edu/abs/2012ApJ...749L..28V} {749, L28}

\bibitem[\protect\citeauthoryear{{Vink{\'o}} et~al.,}{{Vink{\'o}}
  et~al.}{2004}]{2004A&A...427..453V}
{Vink{\'o}} J.,  et~al., 2004, \mn@doi [\aap] {10.1051/0004-6361:20040272},
  \href {http://adsabs.harvard.edu/abs/2004A%26A...427..453V} {427, 453}

\bibitem[\protect\citeauthoryear{{Wheeler} \& {Levreault}}{{Wheeler} \&
  {Levreault}}{1985}]{1985ApJ...294L..17W}
{Wheeler} J.~C.,  {Levreault} R.,  1985, \mn@doi [\apjl] {10.1086/184500},
  \href {http://adsabs.harvard.edu/abs/1985ApJ...294L..17W} {294, L17}

\bibitem[\protect\citeauthoryear{{Woosley} \& {Weaver}}{{Woosley} \&
  {Weaver}}{1986}]{1986ARA&A..24..205W}
{Woosley} S.~E.,  {Weaver} T.~A.,  1986, \mn@doi [\araa]
  {10.1146/annurev.aa.24.090186.001225}, \href
  {http://adsabs.harvard.edu/abs/1986ARA%26A..24..205W} {24, 205}

\bibitem[\protect\citeauthoryear{{Yamanaka} et~al.,}{{Yamanaka}
  et~al.}{2016}]{2016ATel.9124....1Y}
{Yamanaka} M.,  et~al., 2016, The Astronomer's Telegram, \href
  {http://adsabs.harvard.edu/abs/2016ATel.9124....1Y} {9124}

\bibitem[\protect\citeauthoryear{{Yamanaka} et~al.,}{{Yamanaka}
  et~al.}{2017}]{2017ApJ...837....1Y}
{Yamanaka} M.,  et~al., 2017, \mn@doi [\apj] {10.3847/1538-4357/aa5f57}, \href
  {http://adsabs.harvard.edu/abs/2017ApJ...837....1Y} {837, 1}

\bibitem[\protect\citeauthoryear{{Yoon} \& {Langer}}{{Yoon} \&
  {Langer}}{2005}]{2005A&A...443..643Y}
{Yoon} S.-C.,  {Langer} N.,  2005, \mn@doi [\aap] {10.1051/0004-6361:20054030},
  \href {http://adsabs.harvard.edu/abs/2005A%26A...443..643Y} {443, 643}

\bibitem[\protect\citeauthoryear{{Yoon}, {Woosley}  \& {Langer}}{{Yoon}
  et~al.}{2010}]{2010ApJ...725..940Y}
{Yoon} S.-C.,  {Woosley} S.~E.,   {Langer} N.,  2010, \mn@doi [\apj]
  {10.1088/0004-637X/725/1/940}, \href
  {http://adsabs.harvard.edu/abs/2010ApJ...725..940Y} {725, 940}

\bibitem[\protect\citeauthoryear{{Yoshii} et~al.,}{{Yoshii}
  et~al.}{2003}]{2003ApJ...592..467Y}
{Yoshii} Y.,  et~al., 2003, \mn@doi [\apj] {10.1086/375572}, \href
  {http://adsabs.harvard.edu/abs/2003ApJ...592..467Y} {592, 467}

\makeatother
\end{thebibliography}
\label{lastpage}
\end{document}